%% file: 00_main.tex
\documentclass[sigconf,screen]{acmart}

\usepackage{tikz}
\usepackage{amsmath}
\usepackage{etoolbox}
\usepackage{algorithm}
\usepackage{algpseudocode}
\usepackage{booktabs}
\usepackage{tabularx}
\usepackage{graphicx}
\usepackage{xspace}
\usepackage{pifont}
\usepackage{enumitem}

\definecolor{customblue}{HTML}{006ca6}
\definecolor{customgreen}{HTML}{009264}
\definecolor{custombrown}{HTML}{b80d57}
\AtEndPreamble{
\usepackage{hyperref}
 \hypersetup{
  colorlinks = true,
  linkcolor = customblue,
  anchorcolor = purple,
  citecolor = customgreen,
  filecolor = purple,
  urlcolor = black
 }
}

\newcommand{\system}{\textsc{DePra}\xspace}
\newcommand{\one}{({\em i}\/)\xspace}
\newcommand{\two}{({\em ii}\/)\xspace}
\newcommand{\three}{({\em iii}\/)\xspace}
\newcommand{\four}{({\em iv}\/)\xspace}

\setcopyright{none}
\renewcommand\footnotetextcopyrightpermission[1]{}

\begin{document}

\title{Listen to the Voices of Everyday Users: Democratizing Privacy Ratings for Sensitive Data Access in Mobile Apps}

\author{Liu Wang}
\affiliation{%
  \institution{Huazhong University of Science and Technology}
  \city{Wuhan}           
  \country{China}
}

\author{Tianshu Zhou}
\affiliation{%
  \institution{Beijing University of Posts and Telecommunications}
  \city{Beijing}           
  \country{China}
}

\author{Haoyu Wang}
\affiliation{%
  \institution{Huazhong University of Science and Technology}
  \city{Wuhan}
  \country{China}
}

\author{Yi Wang}
\affiliation{%
  \institution{Beijing University of Posts and Telecommunications}
  \city{Beijing}           
  \country{China}
}

\input{01-abstract}

\maketitle
\input{02-intro}

\input{background}

\input{design}

\input{system}

\input{evaluation}

\input{discussion}
\input{relatedwork}
\input{conclusion}


\appendix
\section*{Ethics considerations}
\label{sec:ethics}

This study encompasses two user-related studies: a participatory design activity and user/expert evaluations.
In line with ethical research practices, informed consent was obtained from all participants at the time of recruitment. Prior to their participation, participants were fully informed about the study's objectives, data handling procedures, time commitment, and compensation. 
They were assured that their responses would remain anonymous and that they could withdraw from the study at any point without consequence. 
For the participatory design activity, participants were recruited through community outreach and invited to attend a moderated design session. Each participant received a compensation of ($\approx$)\$10 as a token of appreciation for their time and contributions (which is above the local median hourly wage).
For the end-user evaluations, only those who explicitly agreed to participate in the study were given access to the evaluation platform. Participants who completed the entire questionnaire were fully compensated, irrespective of their responses. The average completion time during the pilot test was 25 minutes, and the reward for participation was set at \$3.8, which is above the minimum wage in the United States. 
As for expert evaluations, all experts were invited based on their background in mobile privacy and policy research. Participation was voluntary, and no monetary incentives were provided.
These two user studies (PD and evaluations) have been reviewed and approved by the Academic Ethics Review Committee of the first author’s institution whose role is equivalent to the Institutional Review Board (IRB) in the United States.

\section*{Data Availability}

The \system platform was developed using a React-based frontend, a Flask-based backend server, and a MySQL database for data storage and retrieval.  
We provide a replication package that contains the source code of \system, the implementation of its key modules (e.g., contextual explanation generation, category-based representative selection), as well as the datasets used in our experiments. The replication package is available at: \url{https://anonymous.4open.science/r/DemocratizePrivacyRating-B7C4/}.

\bibliographystyle{ACM-Reference-Format}
\bibliography{cite}

\input{appendix}

\end{document}

%% file: 01-abstract.tex
\begin{abstract}
Mobile apps frequently request excessive data access, raising significant privacy concerns. While regulations like GDPR emphasize data minimization, they provide limited guidance on concretely defining and enforcing necessary data access. Existing regulatory mechanisms primarily rely on expert-driven audits that face challenges in scalability, neutrality, and alignment with user expectations. In this paper, 
we propose a novel paradigm–democratizing privacy assessment, inspired by prior work on user-centric privacy perceptions–which repositions users as active evaluators in the privacy auditing process, recognizing that user perceptions of data usage play a crucial role in assessing the appropriateness and necessity of data access. To operationalize this paradigm, we introduce \system, a prototype system developed through participatory design, featuring contextual explanation provision, category-based representative selection, an intuitive rating interface, and preference-based rating adjustment. 
We evaluated \system with 200 everyday mobile app users, analyzing how effectively it captures user opinions on sensitive data access, comparing their privacy ratings with expert assessments, and exploring risk preference-based score calibration. Our findings show the feasibility and promise of democratized privacy assessment, highlighting its potential to complement expert auditing and support inclusive privacy evaluation.
\end{abstract}

%% file: 02-intro.tex
\section{Introduction}

Mobile applications (apps) are deeply embedded in people's everyday life, providing users with seamless access to information, e-commerce, social interaction, and health data management.
To enable these functions, apps often request access to personal data and device resources, such as contact lists, geolocation, cameras, and microphones. 
However, many apps exhibit permission-greedy, requesting access beyond what is necessary for their core functionalities~\cite{cybernews,betanews,zdnet,consumerreports}. For example, a Cybernews investigation~\cite{cybernews} of 50 financial apps revealed that over half demanded permissions such as camera, precise geolocation, storage, and contacts, which are not always tightly coupled with the core functions of financial apps. 
Such practices raise concerns about overly broad access to sensitive data by mobile apps.

To combat such overreach, privacy regulations such as the General Data Protection Regulation (GDPR)~\cite{GDPR}, California Consumer Privacy Act (CCPA)~\cite{CCPA}, Personal Information Protection Law of the People’s Republic of China (PIPL)~\cite{PIPL}, introduce principles for governing app data handling. The \textit{data minimization} principle, in particular, requires that data collection and processing be restricted to what is strictly necessary for a given purpose: ``Personal data shall be adequate, relevant and limited to what is necessary in relation to the purposes for which they are processed~\cite{GDPRprinciples}.'' 
However, this principle provides only a broad regulatory framework without concrete, context-specific guidelines for assessing permissible data access. This lack of specificity creates challenges in defining the boundary between ``necessary'' and ``unnecessary'' data access. 
In practice, developers, who often lack privacy expertise (a non-functional requirement), retain significant discretion in making such judgments, which may not always align with user expectations~\cite{zhou2023policycomp}. 
For example, a health app might claim that accessing users' contact lists is ``necessary'' for enabling friend-based challenges or social features. Yet, users may perceive such access as excessive, given that these features are secondary to the app's core functionality of health tracking.
In theory, users could simply refuse such secondary permissions if they consider them unnecessary. In reality, however, the situation is not always seamlessly operationalized: refusal may lead to degraded usability, repeated prompts, or restricted access to certain functions, which, combined with technical complexity and warning fatigue \cite{bongard2022necessary,almuhimedi2015location}, often nudge users toward granting broader permissions than they might otherwise intend without fully considering the implications.
Such situations create opportunities for apps to obtain overly broad access to sensitive data and permissions, thereby exacerbating risks to user privacy.

To ensure the practical enforcement of data minimization, the current approach relies heavily on regulators (e.g., regulatory experts, legal professionals, and privacy auditors) auditing app behaviors and forming legal conclusions. 
While automated tools can assist this process, expert judgment remains indispensable due to the complexity of assessing whether data access is necessary, which typically requires balancing app functionality, regulatory compliance, risk management, and business needs.
However, such an expert-driven evaluation model faces several key challenges: 
(1) \textbf{Scalability and timeliness}: The scarcity of qualified experts and the high labor costs make this approach impractical at scale. With the ever-growing number of apps and their frequent updates, it is challenging for experts to keep pace, causing delays in identifying and addressing privacy issues.
(2) \textbf{Divergent perspectives}: Expert assessments do not always align with end-users’ views. For example, studies~\cite{barth2022understanding,elmimouni2024on} showed that experts and users can prioritize different privacy attributes. When such discrepancies arise, expert-centered assessments may skew developers’ efforts toward satisfying expert criteria, potentially neglecting broader user concerns.
(3) \textbf{Neutrality concerns}: Although experts hold authority over privacy practice evaluations, there is no robust mechanism to guarantee their impartiality. The centralized power raises concerns about fairness and neutrality, especially given the lack of transparent procedures for external oversight.
In summary, reliance on expert-driven evaluations alone is insufficient to establish accurate, equitable, and comprehensive privacy assessment standards.

This paper explores the feasibility of incorporating everyday end-users in app privacy evaluations and considering their perceptions of data usage as a key factor for developers and regulators in assessing compliance with data minimization principles. Several consideratons motivate this idea: (1) \textbf{Scalability and responsiveness}: The large and active user base of mobile apps provides opportunities for closer oversight and quicker feedback. 
Users can review app privacy behaviors in a distributed and parallel manner (e.g., each app has its own user community), prompting timely assessment of privacy practices.
Crowdsourcing of user evaluations could alleviate the scalability and latency challenges of expert-driven assessment models. 
(2) \textbf{Subjectivity in privacy perceptions}: Privacy perceptions are quite subjective; users can have different views on what constitutes necessary data access compared to developers and experts (and even different users have different privacy preferences). Incorporating user feedback brings real-world perspectives that experts may overlook, highlighting areas where evaluations driven solely by theoretical or technical criteria may miss user concerns.
(3) \textbf{Empowering primary stakeholders}: As stakeholders directly affected by privacy risks, users' voices should be valued, and they should play a more active role in shaping privacy standards. This participatory model decentralizes the evaluation process, reducing the disproportionate influence of any single group–be it developers, experts, or regulators. 
By empowering end-users, the system becomes more democratic and better reflects the full spectrum of privacy concerns.

Thus, we propose a novel paradigm–democratizing privacy assessment–that decentralizes the power of privacy rating from minority experts/regulators to the general population of everyday users. 
Inspired by user-centric views of privacy risks, this paradigm positions users as active evaluators who assess the necessity of app data access practices, making privacy assessment more user-relevant and fostering collective sensemaking around privacy issues. 
To realize our paradigm, we adopted a participatory design-informed approach and instantiated \system~\cite{repo}, a prototype web-based platform that allows users to assess and rate app privacy behaviors. The system supports four features: 
(1) \textit{Contextual Explanation Provision}, which presents clear, non-technical explanations of why specific data is accessed and by whom, to support informed user judgments; (2) \textit{Category-Based Representative Selection}, which clusters apps into functional groups and surfaces representative samples to help inform evaluations of similar but unrated apps; (3) \textit{Intuitive Rating Interface}, an interface that presents privacy-related information in an intuitive and accessible manner, enabling users to provide ratings with minimal cognitive burden; and (4) \textit{Preference-Based Rating Adjustment}, a post-processing that adjusts user ratings based on individual risk preferences to improve robustness.
To evaluate the feasibility and capability of \system, we conducted a user study involving 200 participants, who reviewed and rated the necessity of specific data access practices. 
We analyzed the distribution of user rating scores, compared them with expert evaluations, and explored mechanisms for adjusting individual ratings based on privacy risk preferences.
Our findings demonstrated that \system effectively supported user-driven privacy ratings and captured how everyday users perceive the necessity of sensitive data access, enabled systematic comparison with expert evaluations, and calibrated ratings based on individual preferences, offering a scalable and participatory framework that complements expert auditing and supports more inclusive privacy evaluation. 
In summary, we make the following research contributions:
\begin{itemize}
    \item We propose the paradigm of democratizing privacy assessment, transitioning from an expert-centric evaluation model to a participatory framework that values user perspectives in evaluating app data practices.
    \item We design and implement \system, a prototype web-based platform that operationalizes this democratizing privacy assessment paradigm to support everyday users to assess and rate privacy behaviors.
    
    \item We deploy \system in a user study with real-world participants, demonstrating its feasibility and effectiveness in supporting user-driven privacy assessments and highlighting the benefits of the proposed paradigm.
\end{itemize}

%% file: background.tex
\section{Background}

\subsection{Excessive Data Access by Mobile Apps}

Mobile apps frequently require access to sensitive user data to enable various functionalities. To regulate such access, Android and iOS implement permission systems that inform users of intended data use and allow them to grant or deny access~\cite{androidpermission,iospermission}. 
A major risk in this context is over-privilege, where apps request permissions beyond what is necessary for their core functionalities.
It increases the likelihood of sensitive data being collected, misused, or exposed in ways that are not essential for the app’s operation, raising significant privacy concerns.
Moreover, the problem is compounded by the extensive use of third-party libraries,
which developers integrate into apps for functions such as maintenance, analytics, social media, security, and advertising. These third-party libraries inherit the host app's permissions, enabling them to access user data often exceeding what is needed for delivering the app's advertised functionality~\cite{appicaptorblog,backes2016reliable}. 
Users are often left unaware of how their sensitive data is being handled and shared by these embedded components~\cite{tahaei2023stuck,kollnig2021fait}, leaving them vulnerable to potential privacy risks of unintended data exposure.
Consequently, over-privileged permission requests remain a pervasive and pressing issue in mobile ecosystems.

\subsection{Expert-Driven Privacy Auditing}
Traditional privacy auditing is largely conducted by expert teams, including legal scholars, regulatory bodies, and data protection officers. These professionals analyze privacy policies, inspect app codebases, and assess compliance based on legal and technical standards. 
Although this approach delivers legal rigor, some structural limitations have become increasingly evident. For example:

\one \textit{Inspection bottlenecks and case backlogs}: The Irish Council for Civil Liberties (ICCL) showed that 98\% of major cross-border GDPR cases referred to Ireland in 2021 remained unresolved, largely due to the limited technical capacity of Data Protection Authorities (DPAs), with only 9.7\% of their staff are tech specialists~\cite{icclreport}. Additionally, many flagship complaints filed in May 2018 against Big Tech, such as Google and Meta, were still pending four years later, highlighting a persistent enforcement backlog~\cite{wiredreport}.

\two \textit{Sampling bias toward large providers}: Recent GDPR enforcement reports indicated that monetary penalties are disproportionately concentrated on a handful of Big-Tech companies~\cite{gdprreport}. Publicly disclosed cases represent only ``the tip of the iceberg,'' while numerous smaller or ``invisible'' cases are never published by national DPAs. These patterns suggest that enforcement resources and headline sanctions remain concentrated on large, high-profile controllers, leaving the ``long tail'' of smaller app developers with limited oversight.

\three \textit{Misalignment with user expectations}: Expert-led evaluations may not always align with users' real-world expectations or concerns. An obvious example is that expert audits often prioritize legal‐text fidelity, e.g., policy-to-practice consistency and checklist conformance, yet users rarely read privacy policies or base decisions on them~\cite{syrenis,felt2012android,kelley2013privacy}. Instead, user are often more concerned with intuitive and contextual factors such as how, when, and why their data is accessed. This disconnect between formal assessments and user-centric perceptions can lead to inconsistent estimates of privacy risks.

\four \textit{Challenges to regulatory neutrality}: Data protection enforcement could be shaped by bureaucratic conflicts and institutional power struggles, 
particularly when data use intersects with politically or financially sensitive areas~\cite{yesilkagit2011institutional}. For example, Ireland’s DPA,
the lead GDPR authority for Meta due to its EU headquarters in Dublin, has faced repeated criticism from German and French regulators for what they view as lenient or delayed sanctions~\cite{simmonsreport}. 
Such disputes illustrate how national interests and institutional dynamics can influence regulatory outcomes, potentially undermining the consistency and impartiality of data oversight.

%% file: design.tex
\section{Democratized Privacy Rating Paradigm}

Our goal is to develop a democratized privacy rating paradigm that empowers end-users to assess the appropriateness of mobile apps' privacy-related behaviors. 
As an initial step, this study explores the design space and feasibility of such a paradigm by implementing a prototype as its infrastructure, \system (\underline{de}mocratizing \underline{p}rivacy \underline{ra}ting system), 
which enables users to review how apps access sensitive data and resources and make informed assessments about whether such access is justified from their perspective.
We will begin by outlining our design process, then provide an overview of the system, followed by detailed descriptions of its core features. 

\subsection{Infrastructure Design}

\subsubsection{Participatory Design}

To inform the design of \system, we conducted a participatory design (PD) activity.
PD is a well-established creative, collaborative design methodology that begins with users and their experiences and ends with solutions tailored to their specific needs~\cite{aliyu2024participatory,hara2016design}.
In our context, PD allows us to better understand what types of interactive designs, features, and presentation mobile app users desire for future \system. 

Participants were recruited through public advertisements on university mailing lists and local online communities. We asked willing-to-participate respondents to report their basic demographics (gender, age, occupation) and app usage habits (top five frequently used mobile apps and app categories). We also asked several brief questions to gauge their privacy awareness, such as how concerned they were about apps collecting personal data, whether they had ever reviewed or modified app permission settings, and what types of personal data they considered sensitive. 
Respondents were considered in the order they enrolled; when new respondents differed from previous ones in demographics, app usage, or privacy attitudes, they were included to enhance diversity.
After three days of recruitment, we finalized a heterogeneous group of eight participants with diverse demographic backgrounds (4 female, 4 male; aged 20–40; undergraduates, graduates, and office workers), whose frequently used apps covered a broad spectrum including social networking, finance, health, productivity, and gaming. 
None of them had technical expertise in mobile app development or data privacy, which aligns with our target user group for \system.
Although the sample size was modest, we deemed it sufficient for our design aims, as participatory design prioritizes depth of insight and co-creation quality over statistical representativeness.

The PD session lasted approximately 90 minutes and was conducted in a workshop-style format that emphasized open dialogue and mutual learning between participants and facilitators.
The activity was guided by a semi-structured protocol that was developed based on piloting with two graduate volunteers (not included in the final sample), through which we refined timing, improved communication clarity, and adjusted the order of some prompts to better support participants' engagement and reflection during the session.
Overall, the session consisted of three phases:

\noindent \textbf{Phase 1: Introduction and Contextual Framing.}
We began by introducing the purpose of the activity and the goals of the privacy rating system.
We presented participants with concrete examples of apps' sensitive data access behaviors (e.g., “a weather app accessing your location data”), and invited them to reflect on their own experiences.
Participants were asked to discuss what information they would need to judge whether such access was appropriate, and how they typically make such judgments in real-world settings. 
This phase not only elicited user expectations regarding contextual information and privacy decision-making but also established a collaborative environment in which participants' lived experiences defined the design problem space.

\noindent \textbf{Phase 2: Individual Design and Sketching.}
Participants were asked to work individually to imagine and sketch their ideal version of a privacy rating interface, allowing them to express their own values and priorities free from group influence.
To support ideation, we prompted them with open-ended questions such as: \textit{What rating mechanism would feel most natural to you? How much information should be shown, and in what format? How would you like explanations to be presented?} 
We encouraged freehand sketches, annotation on low-fidelity prototypes, and note-taking to express ideas. Participants designed various interface elements, including data visualizations, contextual explanations, and rating widgets (e.g., sliders, Likert scales, emoji reactions).
This phase foregrounded participants’ authorship in shaping potential design directions and encouraged them to externalize personal criteria for evaluating privacy practices.

\noindent \textbf{Phase 3: Sharing and Reflection.}
After sketching, each participant presented their design and rationale to the group. A moderated discussion followed to compare and contrast preferences, such as what contextual information was necessary to support judgments, what kinds of rating schemes felt intuitive, and how much detail could be provided without overwhelming users.
The facilitator guided collective sense-making by encouraging participants to comment on and build upon each other’s ideas.
This reflective exchange fostered mutual learning: participants articulated their reasoning, discovered alternative perspectives, and collectively refined the understanding of which forms of contextual information and presentation approaches better support users in making privacy judgments. 
The discussions revealed recurring themes that directly informed our design directions.

Our approach to participatory design is informed by the Scandinavian tradition that emphasizes democracy, empowerment, and mutual learning~\cite{Gregory2003ScandinavianAT}. 
Rather than treating participants merely as informants, we regarded them as active co-designers whose lived experiences and privacy perspectives shaped the framing of design problems. 
Throughout the workshop, we facilitated opportunities for participants to articulate their values, pondering “appropriate” data practices, and reflect on how their own app usage shaped these judgments. 
This process fostered mutual learning: participants gained greater awareness of privacy implications in everyday apps, while the research team gained a deeper understanding of user perceptions and concerns. 
We recognize that our PD activity represents an early-stage, time-bounded engagement rather than a long-term empowerment process, yet it embodies the core PD ethos of collaborative exploration and reflexivity in the design of privacy-assessment tools.

\subsubsection{Findings from PD}

The session was audio/video-recorded with consent and subsequently transcribed for analysis. Any personally identifiable information was removed to protect participant anonymity during transcription.
We conducted a reflexive thematic analysis~\cite{braun2006thematic} of the transcripts and artifacts (sketches and notes). 
Two authors independently performed open coding on an initial subset of the data, developed a shared codebook, and refined it iteratively through constant comparison. 
Discrepancies were resolved through negotiated agreement. 
We then synthesized first-order codes into higher-level themes using affinity diagramming, which captured recurring design rationales and user expectations. 
Drawing from this analysis, we summarized four main themes that reveal how everyday users perceive and wish to interact with a democratized privacy rating platform.

\begin{itemize}    
\item \textit{Contextual information as indispensable.}
Participants consistently expressed that they could not judge a data access request without knowing \textit{why} the data was needed and \textit{who} was accessing it. 
Moreover, they often grounded their judgments in the core functionality of the app: for instance, location access was considered reasonable for a weather app but not for a flashlight app.
At the same time, participants emphasized that while such contextual details are necessary, too much information at once could be overwhelming.

\item \textit{Accessibility and lightweight rating metaphors.}
Participants indicated that most everyday app users come from diverse backgrounds and varying levels of privacy awareness, and they should not be assumed to have any technical expertise (e.g., how to interpret app code-level behavior). Thus, they expected a user-friendly interface free of technical jargon.
Similarly, they preferred simple rating schemes that felt natural and easy to use, such as Likert-style sliders, thumbs up/down, or emoji-based reactions. They emphasized that the rating process should be quick and minimally effortful, to encourage sustained engagement and scalability.

\item \textit{Selective participation and the need for generalization.} 
Several participants noted that they would likely provide ratings only for apps they personally use or care about. This raised concerns that many niche apps may attract little direct feedback, making it unrealistic to expect sufficient user ratings for every single app. Thus, the system should ideally support strategies for addressing rating sparsity, so that privacy evaluations do not depend on every app receiving equal attention from users.

\item \textit{Diverse privacy attitudes and potential biases.} 
Participants’ discussions revealed that privacy is inherently subjective: individuals vary in their tolerance for data collection and trust in app practices.  
Some expressed concerns that user ratings may inevitably reflect personal biases or outlier opinions, potentially distorting aggregate outcomes if left unmoderated (particularly when the number of ratings is limited). Hence, mechanisms for normalizing or adjusting ratings in light of individual risk preferences are needed to ensure the overall results remain reliable and representative.
\end{itemize}

\subsubsection{Design Goals}

From our PD workshop, we distilled four design goals:

\begin{itemize}
    \item[G1] The system should provide sufficient, context-rich information (but not information overload) regarding each privacy-sensitive behavior in an app to support users in performing informed decision-making.     
    \item[G2] The system should account for selective user participation by enabling representative ratings to be generalized to less-rated or niche apps, ensuring broader coverage despite uneven user engagement.
    \item[G3] The system should be accessible to non-experts (require no specialized technical knowledge), and have an intuitive rating scheme that effectively captures user sentiments and makes users feel easy to use.
    \item[G4] The system should be able to capture individual differences and unique conceptions of what constitutes appropriate data access in mobile apps.
\end{itemize}

%% file: system.tex
\subsection{Overview of \system}

\begin{figure}
    \centering
    \includegraphics[width=0.5\textwidth]{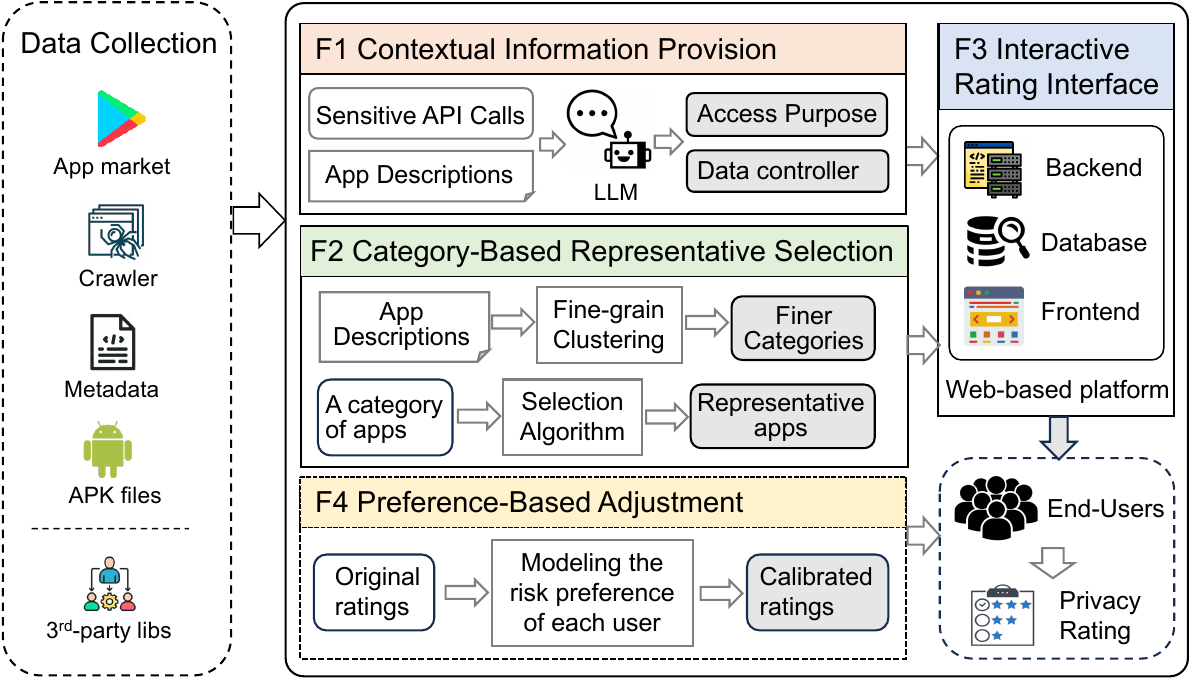}
    \caption{Overview of the \system system.}
    \label{fig:system-overview}
\end{figure}

Enlightened by the above insights, we implemented \system as a web-based platform with a React frontend and a backend server that retrieves app-related information through a connected database.
As illustrated in \autoref{fig:system-overview}, \system involves four core features, each addressing one of the design goals (G1–G4) derived from the participatory design study:
\begin{itemize}   
    \item[F1] \textbf{Contextual Explanation Provision}: To address users’ demand for contextual information, this feature provides insights into apps' sensitive data access behaviors by explaining why specific permissions are requested and who (first-party or third-party) controls the data, paired with the app’s core functionality to support grounded judgments. By situating each data access behavior within its functional context, the feature helps users move beyond vague permission labels and form more informed, context-sensitive opinions.

    \item[F2] \textbf{Category-Based Representative Selection}: To address the challenge of selective user participation, this feature adopts a category-based strategy for potential rating generalization across functionally similar apps. It clusters apps into coherent categories using their descriptions, 
    and within each category selects a minimal yet representative subset of apps that captures the diversity of data access behaviors while providing broad privacy insights that can be extended to other apps in the same category.

    \item[F3] \textbf{Intuitive Rating Interface}: To reflect users’ preference for lightweight rating interactions, the interactive interface is designed to present app behaviors in a clear and accessible way, guiding users to focus on the necessity and acceptability of data access from their own perspective. It minimizes technical jargon, incorporates simple and familiar rating metaphors, and streamlines the evaluation process to enhance usability and engagement.

    \item[F4] \textbf{Preference-Based Rating Adjustment}: To account for diverse privacy attitudes and individual biases, we assessed participants’ risk-taking tendencies through a survey grounded in prospect theory~\cite{kahneman2013prospect}. As an extended feature, this feature operates as a post-processing step after collecting user ratings. The resulting risk profiles (e.g., risk-averse vs. risk-seeking) enabled calibrated adjustments to user ratings, helping to normalize outliers and produce more balanced aggregate results.
    
\end{itemize}

\subsection{Contextual Explanation Provision}
To support informed judgments, \system 
presents the app’s core functionality (\S\ref{sec:description}) and specific purpose explanation along with data controller information (\S\ref{sec:purpose}). 
These elements inform what the app does, why the data is accessed, and who controls it, offering necessary context without causing information overload.

We note that obtaining reliable, non-technical explanations of app data access behaviors is itself a non-trivial challenge, and a substantial line of prior work has sought to address this through techniques such as static/dynamic analysis, privacy policy mining, and GUI-based inference~\cite{yang2022describectx,wang2017understanding,zhou2023policycomp,wang2025big}.
The focus of this work is not on solving this data collection problem, but on demonstrating the value and feasibility of the democratized privacy assessment paradigm once such explanations are available. 
For this study, all deployed items were verified by the research team prior to deployment to ensure reliability.
We also note that in real-world deployments, such explanations need not always be derived through external analysis. Developers themselves could possess knowledge of why their apps access specific data; they could supply these purpose descriptions to the platform, enabling users to evaluate the acceptability of the developer's own stated data practices. This would represent a particularly natural instantiation of the paradigm, where user ratings serve as a form of participatory validation of developer-declared data use rationale.

\subsubsection{App Description}
\label{sec:description}
To understand an app's functionality, we leveraged app descriptions, i.e., the natural language descriptions in app marketplaces serving as proxies for an app's advertised functionality. Previous studies~\cite{gorla2014checking,zhou2023policycomp} have demonstrated that analyzing app descriptions effectively identifies core functionalities.
We thus chose app descriptions as well as the provided screenshots from the official app market to help users understand the basic context of app usage.
These materials were collected using the open-source web crawler~\cite{gpscraper}, stored in our database, and integrated into \system's interface, where they are displayed on demand to support user evaluations.

\begin{figure*}
    \centering
    \includegraphics[width=0.9\textwidth]{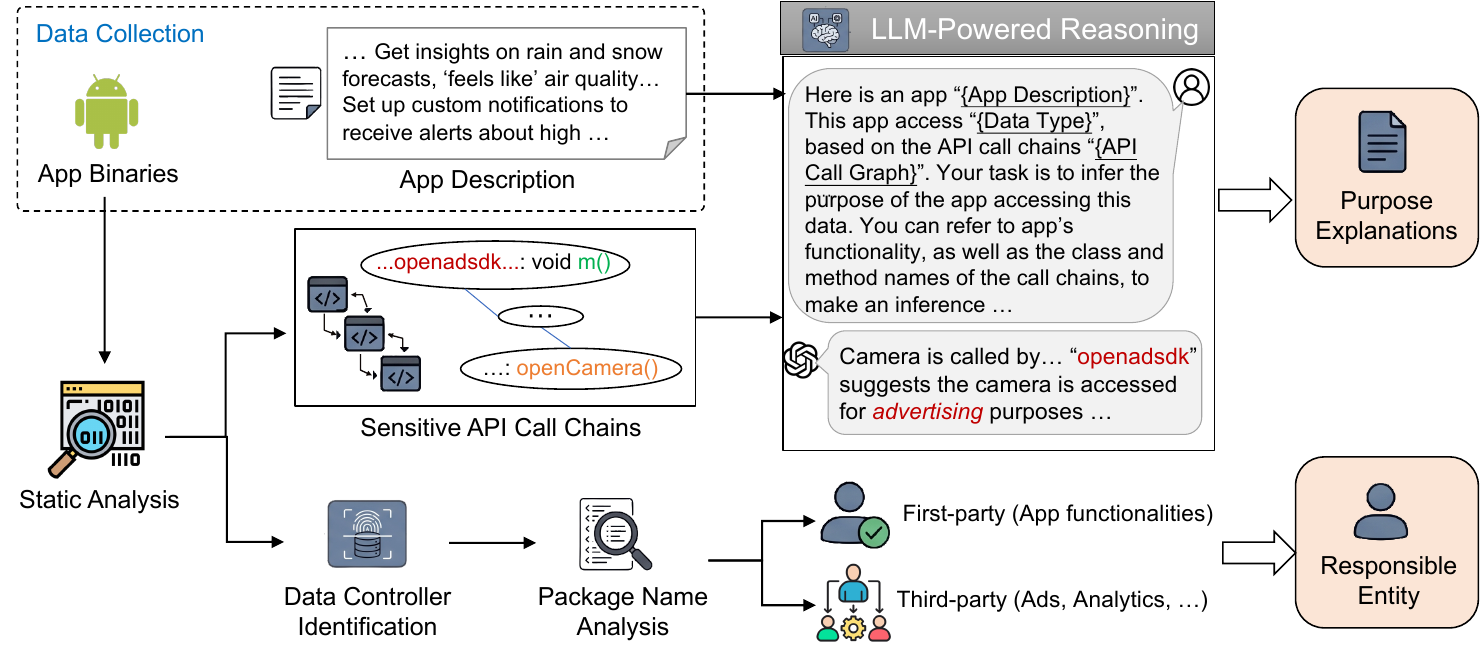}
    \caption{Sensitive behavior and purpose inference for an app.}
    \label{fig:purpose-inference}
\end{figure*}

\subsubsection{Purpose Description and Data Controller}
\label{sec:purpose}
Besides app descriptions, it is equally crucial to explain why sensitive data is accessed and who controls it.
For purpose description, prior studies have demonstrated thatsensitive API call chains and app metadata can effectively support inference of the purpose behind app behaviors~\cite{yang2022describectx,wang2017understanding}. 
Building on this line of work, we combine sensitive API call chains with app store descriptions and use large language models (LLMs) to generate purpose explanations, as shown in \autoref{fig:purpose-inference}.
Specifically, we implemented a sensitive-API tracker using static analysis on app binaries to extract sensitive API invocation traces and their corresponding call chains. 
Here we focused on high-impact sensitive resources (following prior research~\cite{lin2012expectation}), including \textit{calendars, call logs, contacts, location, device IDs, body sensors, SMS, storage, camera, and microphone}. 
The list of our concerned data and resources is presented in Appendix~\ref{sec:tables} Table~\ref{tab:data_permissions}.
We retrieved the corresponding API call stack traces and employed LLM-powered reasoning to infer the most likely purpose of each data request.
We designed structured prompts that guide the LLM (gpt-4o) in analyzing sensitive API call stacks in conjunction with app descriptions, providing contextual cues to determine the most probable purpose of each data request.

Beyond explaining the purpose of data access, it is equally important to convey who is responsible for it. 
The distinction between first-party (app-internal) and third-party (external service) data access is crucial, as it significantly influences user privacy perceptions and risk assessments.
To address this, we categorize data controllers based on whether access is initiated by the app itself (first-party) or by an embedded third-party service (e.g., advertisers, analytics providers). Following prior work \cite{lin2014modeling}, we group third-party services into ten distinct categories (as shown in Appendix~\ref{sec:tables} Table~\ref{tab:sdk_types}). 
Since API call chains trace the sequence of function calls leading to sensitive API invocations, we analyze package names within the call chain to determine the responsible data controller. Our classification strategy is straightforward: If the sensitive API invocation occurs within a third-party package, we classify the data access as third-party usage; otherwise, it is considered app-internal. 
We maintain a database of third-party library package names collected from open-source academic artifacts, which facilitates the classification\footnote{The database is continually updated to include newly identified third-party services, aiming to provide as comprehensive coverage as possible.}. 
We acknowledge that this binary classification is a practical simplification: as noted in prior work~\cite{nguyen2021share}, when a developer voluntarily integrates an analytics or advertising SDK, the developer remains the data controller under GDPR, since it is the developer's own decision to collect data via that SDK. Our classification, which marks such cases as third-party based on code origin, does not accurately capture this controller relationship. It is intended to convey an intuitive signal to users about whether external code is involved in data processing, rather than a precise legal determination of controller status. 

Overall, the automated pipeline described in this section serves as a proof-of-concept implementation; we do not claim it to be a production-ready or comprehensively evaluated technical solution. The current pipeline draws on a specific subset of available information sources, namely, static API analysis and app store descriptions. Other sources that could enrich contextual explanations include the Data Safety Section on Google Play, GUI-level rationale messages that precede permission requests, privacy policy text, and opt-in/opt-out dialogues provided by privacy-configurable SDKs. Incorporating these complementary sources would likely yield more complete and accurate purpose descriptions, and is an important direction for future work.

\subsection{Category-Based Representative Selection}
To address the potential imbalance of user-driven privacy ratings,
\system adopts a category-based strategy that increases the utility of collected ratings. We first cluster apps into coherent functional groups (\S\ref{sec:clustering}) and then select a small set of representative apps from each group (\S\ref{sec:appselection}). This approach helps evaluations remain broadly informative and transferable within the same category.

\subsubsection{App Functionality Clustering}
\label{sec:clustering}
\begin{figure*}
    \centering
    \includegraphics[width=\textwidth]{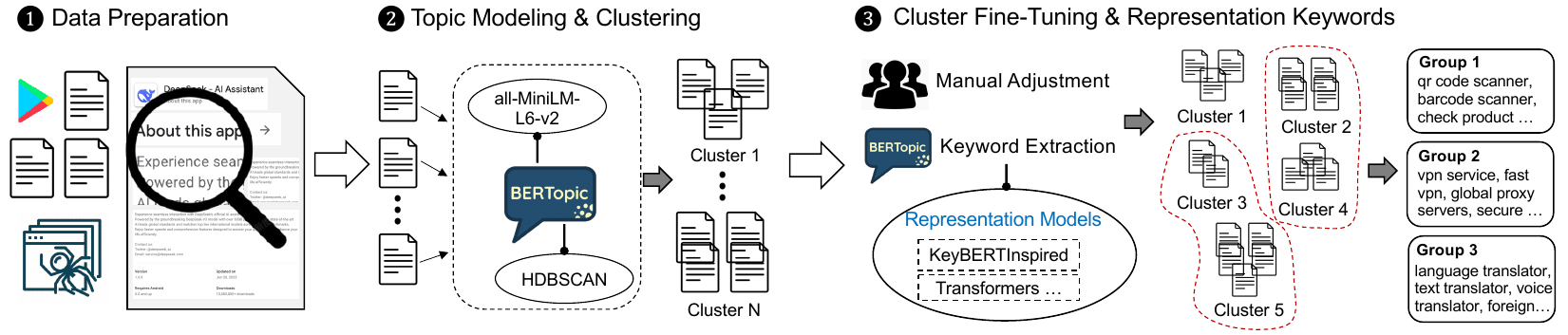}
    \caption{App description-based clustering for each broad (Google Play Store provided) category.}
    \label{fig:clustering}
\end{figure*}

We first leveraged app descriptions to cluster apps with similar functionality and privacy requirements.
Major app marketplaces such as Google Play and App Store already provide category labels like \textit{Business}, \textit{Education}, and \textit{Entertainment} to facilitate user searches and discovery. However, we cannot directly use these predefined labels as our categorization since they are often too broad for precise privacy assessments. For example, the \textit{Tools} category in Google Play and the \textit{Utilities} category in the App Store include vastly different apps, such as flashlight apps and calculator apps, which serve distinct purposes and have varying privacy implications. Nevertheless, these labels serve as a useful starting point for narrowing the search space and grouping potentially related apps.
Specifically, we introduced a refined categorization framework that subdivided marketplace categories into granular subgroups. Each subcategory consisted of apps that exhibit similar functionalities and privacy requirements, enabling more precise and meaningful privacy evaluations.
\autoref{fig:clustering} illustrates our three-step clustering methodology. First, we compiled a comprehensive dataset from Google Play, extracting app descriptions for input. Then we applied BERTopic~\cite{BERTopic}, a state-of-the-art topic modeling technique, to generate semantic embeddings of app descriptions. The output clusters then underwent manual refinement, where we adjusted categories to ensure semantic coherence and privacy relevance. A set of representative keywords would be generated for each cluster, summarizing their functional themes. 
We elaborate on the steps in the following.

\noindent \textbf{(1) Preparing App Data.}
 
To collect apps, we relied on AndroZoo~\cite{allix2016androzoo}, a growing repository of Android apps collected from various sources. From this dataset, we extracted a diverse collection of apps sourced from Google Play, ensuring broad coverage of different functionality types. For each app, we retrieved its detailed metadata using the web crawler~\cite{gpscraper},
and then pre-processed the app descriptions by removing non-English entries and short ones (fewer than 30 words). The resulting corpus of descriptions served as the primary input for our clustering process.

\noindent \textbf{(2) Topic Modeling and Clustering.}
Various topic modeling techniques have been explored in prior research~\cite{egger2022topic,axelborn2023topic,khodeir2024efficient}, 
with BERTopic demonstrating strong performance. Empirical study~\cite{bu2023software} has shown that BERTopic is particularly effective for clustering the description texts of application software. As a state-of-the-art topic modeling approach, BERTopic utilizes pre-trained BERT models to generate contextual text embeddings, which transform textual descriptions into continuous vector representations. These representations are then clustered using HDBSCAN,  
while class-based term frequency-inverse document frequency (c-TF-IDF) is applied to enhance term representation within clusters. To further improve computational efficiency, Uniform Manifold Approximation and Projection (UMAP) is used to reduce the dimensionality of the embeddings before clustering. Since BERTopic operates in an unsupervised manner, it does not require predefined topic labels or training data, making it highly adaptable for large-scale text analysis.

By leveraging BERTopic, we efficiently processed and analyzed large volumes of app descriptions, constructing a structured topic space that facilitates the categorization of apps based on functionality.
We applied BERTopic within individual major marketplace categories. The app descriptions from a specific category were input into BERTopic, 
which formed functionally coherent subgroups based on the density distribution of data points in the embedding space. 
To ensure a granular topic discovery, we set a small \texttt{min\_topic\_size} parameter, defined as the total number of apps in a category divided by 20. This parameter assumes each category can be subdivided into at most 20 more specific subcategories, a deliberate over-segmentation of the data, aiming to minimize the risk of missing niche functionalities within a category and allow for a comprehensive and fine-grained clustering.

\noindent\textbf{(3) Cluster Fine-Tuning and Representation Keywords.}
BERTopic offers multiple strategies for generating topic representations, providing diverse perspectives on the extracted clusters. After conducting preliminary evaluations, we integrated KeyBERT and LLM-based models as primary representation models within BERTopic to enhance the interpretability of clusters. Specifically, we employed \texttt{KeyBERTInspired()} and OpenAI's GPT-based models to extract semantically meaningful keywords that effectively characterize each cluster's functional theme.

Following the clustering process, we conducted a manual review to further refine and validate the results. This involved analyzing the extracted keywords and visualizing topic distributions to identify redundancies and potential overlaps. 
Our manual refinement process adhered to a key principle: \textit{we do not aim for excessive functional granularity; we differentiate categories only when distinct functionalities are likely to imply different privacy requirements–that is, the typical and justifiable data-access needs associated with an app’s core functionality}.
For instance, in the \textit{Events} category, although there are multiple event types (e.g., conferences, concerts, sports events), they generally do not differ significantly in terms of privacy-related behaviors from a user perspective (e.g., they all involve event management functionalities). Therefore, we did not separate them into finer subcategories. In contrast, within the Tools category, different types of tools often have distinct privacy implications–for example, a flashlight app, a calculator, and a VPN service, each handles user data in fundamentally different ways. In such cases, we created finer distinctions to better capture the variations in privacy needs across apps.
When multiple clusters exhibited similar topics with overlapping privacy requirements (e.g., conferences, concerts, etc.), we manually merged related clusters using BERTopic's built-in \texttt{merge\_topics} function. This consolidation enhanced the coherence and interpretability of the generated topics, ensuring that each final cluster represented a distinct and meaningful functional grouping. 
Finally, each finalized cluster was described by a set of representation keywords derived from representation models, providing a concise yet informative summary of its functional characteristics.

\subsubsection{Representative App Selection}
\label{sec:appselection}
We developed a selection strategy aimed at identifying a set of representative apps within each functionality-based category. 
The selection criteria prioritized data access diversity, and we also considered the number of installations, giving preference to widely used apps when their data access behaviors are comparable. However, it is computationally expensive and time-consuming to comprehensively identify data access behaviors for each app. To streamline the process, we leveraged the permission information disclosed on the Google Play page as alternatives.
Google Play Store provides a standardized section within the ``About this app'' panel, where permissions required by an app (e.g., camera, location, storage) are explicitly listed. This section serves as a critical resource for users to evaluate privacy implications prior to app installation, and its standardized presentation ensures consistent visibility across all apps. 
These permission requests serve as an explicit indicator of potential personal data access, making the permission list a useful proxy for assessing an app’s privacy behavior. While the permission list alone does not reveal how the data is actually used, it offers an initial reference point to mirror personal data access practices and aids in selecting representative apps.

To determine the minimal set of representative apps, we employed a greedy algorithm that prioritizes coverage of unique permissions while considering install counts as a secondary factor. The algorithm operates as follows (Algorithm 1): It begins by constructing the universal set of permissions by aggregating all permissions across the given set of apps. A set of uncovered permissions is maintained and initialized as the full set of permissions that need to be covered. An initially empty set of selected apps is used to store the final representative apps. In each iteration, the algorithm selects the app that covers the largest number of uncovered permissions. If multiple apps cover the same number of permissions, the app with the highest install count is chosen to prioritize more widely used apps. Once an app is selected, it is added to the representative set, and the permissions it covers are removed from the uncovered set. This process repeats until all permissions in the universal set have been covered by the selected apps. The algorithm results in a minimal subset of apps that ensures comprehensive permission coverage. By employing this approach, the selected representative apps not only achieve comprehensive permission coverage but also favor commonly used apps, increasing their relevance in real-world scenarios.

\subsection{Intuitive Rating Interface}
To support user evaluations of app data access behaviors, we designed an interactive web interface that 
is structured to make rating tasks focused and streamlined (\S\ref{sec:ratingflow}) and applies user-centered design choices to make it accessible for a wide range of users (\S\ref{sec:usercentered}).
These design decisions were directly informed by the PD findings: participants stressed that the system should require no specialized technical knowledge, and that the rating scheme should feel natural and minimally effortful to encourage sustained engagement.

\begin{algorithm}
    \caption{Selecting Minimal Representative Apps}
    \label{alg:select_representative_apps}
    \small
    \begin{algorithmic}[1]
        \Require A set of apps $A$, where each app $a \in A$ has:
        \Statex \quad - a set of permissions $P_a$
        \Statex \quad - a filename $f_a$
        \Require A dictionary $Installs$ mapping each filename $f_a$ to its install count.
        \Ensure A minimal set of representative apps $S \subseteq A$ that covers all permissions.
        
        \State $all\_permissions \gets \bigcup P_a$ for all $a \in A$
        \State $remaining\_permissions \gets all\_permissions$
        \State $S \gets \emptyset$
        
        \While{$remaining\_permissions \neq \emptyset$}
            \State $best\_app \gets \text{None}$
            \State $best\_coverage \gets \emptyset$
            \State $highest\_installs \gets -1$
            
            \For{each app $a \in A$}
                \State $coverage \gets P_a \cap remaining\_permissions$
                \If{$|coverage| > |best\_coverage|$ \textbf{or} 
                     ($|coverage| = |best\_coverage|$ \textbf{and} $Installs[f_a] > highest\_installs$)}
                    \State $best\_app \gets a$
                    \State $best\_coverage \gets coverage$
                    \State $highest\_installs \gets Installs[f_a]$
                \EndIf
            \EndFor
            
            \State $S \gets S \cup \{f_{best\_app}\}$
            \State $remaining\_permissions \gets remaining\_permissions \setminus best\_coverage$
        \EndWhile
        
        \State \Return $S$
    \end{algorithmic}
\end{algorithm}

\subsubsection{System Interface and Rating Workflow}
\label{sec:ratingflow}
\begin{figure}
    \centering
    \includegraphics[width=0.5\textwidth]{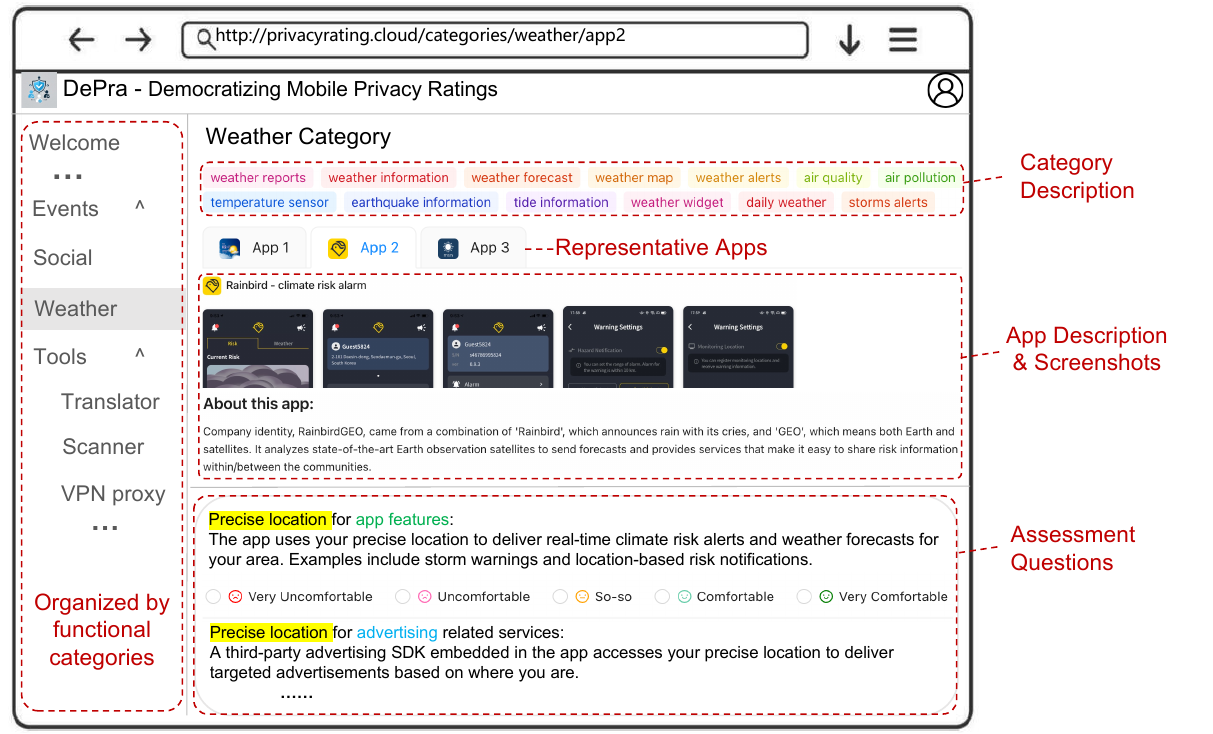}
    \caption{The \system user evaluation interface, comprising: (1) a category-based navigation panel on the left; (2) a category description and representative app selection at the top; (3) app description and screenshots in the middle; and (4) assessment questions at the bottom, each presenting a contextual explanation of a specific data access behavior followed by a comfort rating scale.}
    \label{fig:interface}
\end{figure}

The user interaction is facilitated through a well-organized web interface that aims to streamline the evaluation process.
As illustrated in \autoref{fig:interface}, the interface is organized by functional app categories (e.g., Weather, Tools-Transltor Tools). Within each category, a set of representative apps is featured. 
App-specific information, including screenshots and market descriptions, is presented on demand: users can expand it to understand an app’s core functionality before rating, but it does not clutter the primary rating view. This layered disclosure strategy responds directly to the PD finding that participants wanted sufficient context without being overwhelmed, keeping essential rating information in focus while making broader context accessible.
We designed each Human Intelligence Task (HIT) as a set of focused questions regarding a specific data access request and its stated purpose. 
For each question, users are presented with a two-tier format: a concise header identifying the data type, purpose, and data controller at a glance (e.g., ``Precise location for app features'' or ``Precise location for advertising related services''), followed by a brief natural language explanation beneath it offering additional context for those who wish to read further before rating. This two-tier structure balances immediacy with clarity: the header enables quick comprehension, while the explanation supports more informed judgment, reflecting the PD insight that users want to make quick judgments without being forced to read dense descriptions, yet still have access to richer context when needed.
Users rate their comfort level with each behavior using a 5-point Likert scale ranging from very comfortable (+2) to very uncomfortable (-2). During the PD workshop, participants proposed various rating formats including thumbs up/down and emoji-based reactions. 
We opted for a symmetric numeric Likert scale because it captures the continuous gradient of acceptability more expressively than binary options, and the signed range naturally distinguishes two directions of user judgment: positive scores indicate acceptance, negative scores indicate privacy concern, with zero serving as a neutral baseline. This bidirectional structure makes the scale intuitive for users to interpret and apply, without requiring them to reason about abstract numerical anchors.

\subsubsection{User-Centered Design Considerations}
\label{sec:usercentered}
To enhance accessibility and usability for a diverse audience, we prioritized minimizing cognitive load and avoiding technical jargon–principles well-established in the design of effective privacy notices~\cite{schaub2015design}. Concretely, we used plain, user-friendly terminology to describe data types: instead of ``coarse (Network) location,'' users see ``approximate location''; instead of raw permission names, they see human-readable labels. For unavoidable technical terms (e.g., device-specific identifiers, IMEI), we added inline tooltips that surface a brief explanation on hover, allowing curious users to learn more without burdening those who do not need the detail. Together, these choices ensure that non-expert users can engage meaningfully with privacy assessments without requiring prior knowledge of Android permission semantics or privacy regulations.

\subsection{Preference-Based Rating Adjustment}

To mitigate the impact of individual differences in privacy attitudes and risk tolerance, we introduce a post-processing adjustment mechanism that models users’ privacy risk preferences (\S\ref{sec:theory}) and recalibrates their ratings accordingly (\S\ref{sec:adjustmethod}).
As an extended feature of \system applied after collecting raw ratings, this mechanism aims to help aggregate results better reflect general user sentiment rather than being skewed by extreme or biased evaluations.

\subsubsection{Theoretical Foundation}
\label{sec:theory}
Our current approach builds on \textit{prospect theory} \cite{kahneman2013prospect}, a well-established framework in behavioral economics that characterizes how individuals make decisions under conditions of risk and uncertainty. Prospect theory has been widely adopted in prior research on privacy decision-making~\cite{liao2019prospect,qu2019towards,keith2014longitudinal}, where it has been shown to capture users’ attitudes toward potential losses of personal data and their willingness to trade privacy for utility. 
Inspired by this line of work, we use prospect theory to guide the measurement and categorization of user risk attitudes in privacy ratings.

Thus, we can design a set of scenarios to measure users' risk-taking behavior in contexts involving both potential gains and losses. Such scenarios reflect diverse decision-making environments, e.g., considering monetary incentives, penalties, or potential utility trade-offs. Users’ responses provide insights into their general tendencies toward risk, offering a basis for interpreting their responses and identifying potential biases in their evaluations.
Based on user responses for the scenarios, we can classify them into three basic groups: 1) \textit{Risk-averse}: individuals who generally exhibit conservative tendencies, preferring to avoid potential losses and showing reluctance to accept privacy risks. 2) \textit{Risk-seeking}: individuals who are more tolerant of uncertainty and willing to accept or even favor risky data practices. 3) \textit{Risk-neutral}: individuals whose responses fall in between, showing no strong inclination toward either avoiding or embracing risk, and typically exhibiting moderate or balanced attitudes.
This classification provides the foundation for calibrating privacy ratings in a manner that accounts for systematic differences in user attitudes toward risk.
 
\subsubsection{Adjustment Mechanism}
\label{sec:adjustmethod}
 
A core insight from prospect theory is that risk-averse individuals weigh potential losses more heavily than equivalent gains~\cite{kahneman2013prospect}. Prior work has found that individual risk aversion carries over into privacy decisions: more risk-averse individuals tend to be less willing to incur privacy risks and exhibit more cautious attitudes toward data disclosure~\cite{frik2020measure}. Consistent with this, our underlying assumption is that risk-averse individuals tend to assign lower privacy ratings (i.e., greater discomfort with data access), while risk-seeking individuals are more inclined to assign higher ratings.
Left unadjusted, these tendencies may skew aggregate results toward the extremes.
To mitigate this tendency, we introduce a \textit{risk aversion factor} $\lambda$ ($0 < \lambda < 1$) and an \textit{adjustment coefficient} $\delta$ ($\delta > 0$), which together modulate ratings in proportion to participants’ risk orientations. 
Formally, let $r$ denote a participant’s original rating. The adjusted rating $r'$ is computed as:
\begin{itemize}[leftmargin=*, nosep]
    \item For risk-averse users, scores are increased proportionally to their aversion level:

\begin{equation}
    r' = r + \lambda \delta
\end{equation}

    \item For risk-seeking users, scores are decreased accordingly:

\begin{equation}
    r' = r - (1-\lambda) \delta
\end{equation}
    
\end{itemize}

This adjustment increases scores from risk-averse individuals (who may otherwise systematically underrate) and decreases scores from risk-seeking individuals (who may otherwise overrate), resulting in a more balanced and representative aggregation.
The parameters $\lambda$ and $\delta$ should be informed by the distributional characteristics of the participant group, which helps ensure that adjustments reflect the overall user population and yield a context-aware correction rather than arbitrary normalization.

%% file: evaluation.tex
\section{Evaluation}
In this section, we evaluate whether \system supports effective user participation in privacy ratings and explore its broader usefulness for privacy assessment practices. 
Our evaluation is guided by the following research questions:
\begin{itemize}
    \item \textbf{RQ1:} How effective is \system in democratizing privacy ratings by capturing everyday users' opinions on sensitive data access?
    \item \textbf{RQ2:} How do user-generated privacy ratings compare with expert assessments?
    \item \textbf{RQ3:} Can we account for individual user biases and different risk preferences to improve the robustness of user‑driven privacy ratings?

\end{itemize}

\subsection{Experimental Setup}
\subsubsection{Participants}
We employed the Prolific~\cite{Prolific} crowdsourcing platform to recruit end-user participants for our experiment.  
Prolific offers access to a large and internationally diverse participant pool with built-in quality-control and fair-compensation mechanisms, making it well-suited for behavioral and HCI studies requiring reliable online data collection~\cite{10.1145/3706598.3713251,10.1145/3706598.3713783,kwesi2025exploring}. 
Our participant screening was designed to be diverse, without imposing specific demographic restrictions, in order to capture a wide range of user perspectives on privacy across different backgrounds. 
A total of 200 participants were recruited,
with the demographic characteristics outlining in \autoref{tab:demographics}. The sample comprises a diverse mix of genders, ages, ethnicities, and nationalities, with 118 females, 82 males, and representation from 26 different countries. This diversity provides a broad perspective on privacy perceptions across different backgrounds.

In addition to end-users, we also recruited four domain experts from the authors' professional networks: two privacy researchers from academia and two compliance professionals from regulatory agencies, with an average of six years of experience in data privacy, mobile security, or regulatory compliance. Their ratings provide a reference point for comparing expert-driven assessments with user-generated ratings in later analyses.

\begin{table}[htbp]
    \centering
    \caption{Demographics of the 200 participants.}
    \resizebox{\linewidth}{!}{
    \begin{tabular}{lc}
        \toprule
        Gender & Female / male: 118 / 82 \\
        Age & Min / Median / Max / Mean: 18 / 32.8 / 60 / 30\\
        Ethnicity simplified  & Asian / Black / Mixed / Other / White: 3 / 146 / 6 / 3 / 42 \\
        \# of Nationality & 26 (US, UK, South Africa, Zimbabwe, Spain, Poland, ...)\\
        Language & English / Other: 169 / 31\\
        \bottomrule
    \end{tabular}
    }
    \label{tab:demographics}
\end{table}

\subsubsection{Privacy Rating Tasks}
In this experiment, we asked participants to evaluate their perceived willingness to accept sensitive data access requests by mobile apps using \system. 
We randomly selected six functional clusters for experiment: \textit{Weather}, \textit{Social}, and \textit{Events}, which are marketplace-defined categories that do not require further subdivision\footnote{These categories exhibit sufficient functional specificity, meaning that their core functionalities and data access requirements are relatively consistent. For example, weather apps primarily provide forecasts and related services, social apps facilitate interactions and networking, and event apps manage event-related activities.}, along with three subcategories from the \textit{Tools} category (\textit{translator tools}, \textit{scanner tools}, and \textit{proxy VPN tools}).
We then identified 1-3 representative apps per category that collectively encompass the permission usage scenarios observed in their respective groups (totally 13 apps). For each selected app, the app description, every instance of sensitive data access behavior, and the corresponding purpose explanation were presented on the rating interface. 
In total, 78 privacy assessment questions were presented for these categories.
A linear navigation flow was implemented to guide users to sequentially evaluate each app, ensuring structured and systematic feedback collection.
Prior to deployment, all 78 items (including both the purpose explanations and the first/third-party controller classifications) were individually reviewed by members of the research team, who cross-checked each item against the app's store description, privacy policy, and observed API behavior, with corrections applied where necessary. The validity of our findings therefore rests on the quality of these verified items rather than on the end-to-end accuracy of the automated pipeline.

To ensure valid responses, we embedded several attention-check questions in the task. These questions, such as asking participants to identify the app category they were assessing, can detect inattentive or low-effort responses (e.g., random clicking without reading the questions). Participants who failed these checks were flagged as inattentive, and their responses were excluded from the final analysis.

Note that, before launching the main experiment, we conducted a pilot test with three volunteers (two had participated in the earlier participatory design study, and one was new to the project) to operationalize these privacy rating tasks. 
The pilot aimed to assess the clarity of task wording, confirm the appropriateness of questions, and estimate overall completion time. 
Based on pilot feedback, we rephrased several permission-purpose explanations to avoid ambiguity, simplified the layout of a few longer app descriptions, and slightly adjusted the placement of attention checks to maintain engagement without interrupting task flow. 
The average completion time of pilot tests was 25 minutes, which informed our compensation rate on Prolific to set at \$3.8 per participant, slightly above the minimum hourly wage in the United States.
These iterative refinements served as a formative evaluation of the interface and task design, ensuring accessibility and clarity prior to full deployment.

We also note that ratings in this study are based solely on the contextual information presented by \system; participants may not have had prior first-hand experience with the rated apps. This mirrors the conditions under which users encounter privacy information in practice (e.g., at install time or when reviewing a privacy notice), and is consistent with prior usable privacy research that studies user responses to structured privacy disclosures~\cite{schaub2015design,kelley2013privacy}. The trade-off is that judgments may not fully reflect the nuanced understanding that comes from extended app interaction.

\subsubsection{Participants' Perception of Privacy}
We also included a post-study survey (detailed in Appendix~\ref{sec:survey}) to understand these participants' privacy awareness and attitudes. 
The survey was administered after the main rating tasks to avoid potential priming effects that might bias participants’ in-task judgments about app data access scenarios.
Part of the survey asked about their sensitivity to different data types, their awareness of privacy policies, and their app behavior choices when confronted with potential privacy risks, which could provide valuable context for understanding participants' privacy sense. 
To ease completion and ensure comparability, the survey primarily consisted of single- or multiple-choice questions, with a few optional open-ended fields allowing participants to elaborate if desired. 
Quantitative analysis, complemented by limited qualitative inspection, showed that participants overall demonstrated a moderate level of privacy awareness among participants. Specifically, 
104/70/26 of them reported that they often/sometimes/rarely carefully read privacy policies when installing apps. 
146/49/5 claimed they were very/somewhat/not familiar with setting app permissions. When asked to choose between free apps requiring extensive permissions and paid apps that do not collect personal data, responses were almost evenly split (105 vs. 95), highlighting diverse privacy preferences among participants.
These findings suggest that participants exhibit a broad spectrum of privacy attitudes and behaviors, 
offering useful diversity for examining differences in privacy ratings, though not a demographically representative population sample.
The other part of the survey focused on participants’ general attitudes toward risk in decision-making contexts, which were later used to contextualize their privacy risk preferences and to calibrate our risk preference–based adjustment model (\S\ref{sec:rq3}).

The survey also included an open-ended question asking participants to share their general thoughts on mobile privacy protection. Responses were analyzed thematically~\cite{braun2006thematic}, revealing several recurring themes that further characterize the participant sample.
The most prevalent theme was a strong recognition of privacy as important, with many participants expressing heightened concern about personal data exposure (e.g., \textit{``Mobile privacy protection is increasingly vital in today’s digital landscape, where personal data is frequently at risk’’}).
A second theme concerned skepticism toward existing protection mechanisms: several participants expressed uncertainty about whether current safeguards actually work in practice (e.g., \textit{``I don’t know if my privacy is really protected or not, I don’t usually trust them much’’; ``there’s no way of knowing if they keeping end of the bargain on the privacy policies’’}).
A third theme highlighted the cognitive burden of managing privacy: participants noted that complex policies and permission interfaces can be overwhelming and discouraging (e.g., \textit{``It can be intimidating to go through all the policies and would discourage a lot’’; ``its hard to keep up with the permissions apps have’’}).
Finally, a subset of participants described active permission-management behaviors, such as reviewing permissions before installation and revoking unnecessary ones (e.g., \textit{``I review app permissions carefully before installing and revoke permissions for apps that don’t need them’’}), while others acknowledged limited capacity to do so consistently.
Taken together, these qualitative responses indicate that participants held genuine and varied privacy concerns, ranging from principled data minimization expectations to pragmatic acceptance of common industry practices, providing a meaningful basis for the privacy assessments that follow.

\subsection{RQ1: Capability of \system}

\subsubsection{Method}
We evaluated the feasibility and effectiveness of \system in supporting the democratization of privacy ratings by inviting end-users to perform privacy assessments using \system.
We operationalize effectiveness as \system's ability to enable everyday users to produce informative privacy ratings, i.e., whether the collected ratings capture meaningful variation in user opinions across different apps and data access behaviors. 
We published this study on Prolific, where participants were provided with a brief overview of the study objectives, task requirements, and a direct link to the evaluation platform. Individuals who chose to participate and gave consent would be redirected to the platform via the link.
Upon accessing the platform, participants were greeted with a \textit{Welcome} page that provided step-by-step instructions on completing the evaluation. The instructions covered key aspects such as platform navigation, evaluation criteria, and responding to privacy-related questions. 
After participants completed their evaluations and submitted responses, we conducted a review process to verify that all required questions were answered. Valid and complete submissions were approved, and participants received compensation accordingly. In our experiment, only one submission was excluded from approval due to incomplete responses to essential questions. To maintain the intended sample size, we subsequently reopened recruitment on Prolific to replace the excluded submission with a new participant.

\subsubsection{Results}
All participants completed the assessment tasks over the course of two days. After excluding three responses that failed quality control checks, we obtained a total of 15,366 valid responses from 197 participants. We next analyze their rating scores at each behavior level and app level.
We use violin plots for visualization to illustrate the full distribution of ratings, highlighting where responses are concentrated and whether opinions are polarized.

\begin{figure}
    \centering
    \includegraphics[width=0.5\textwidth]{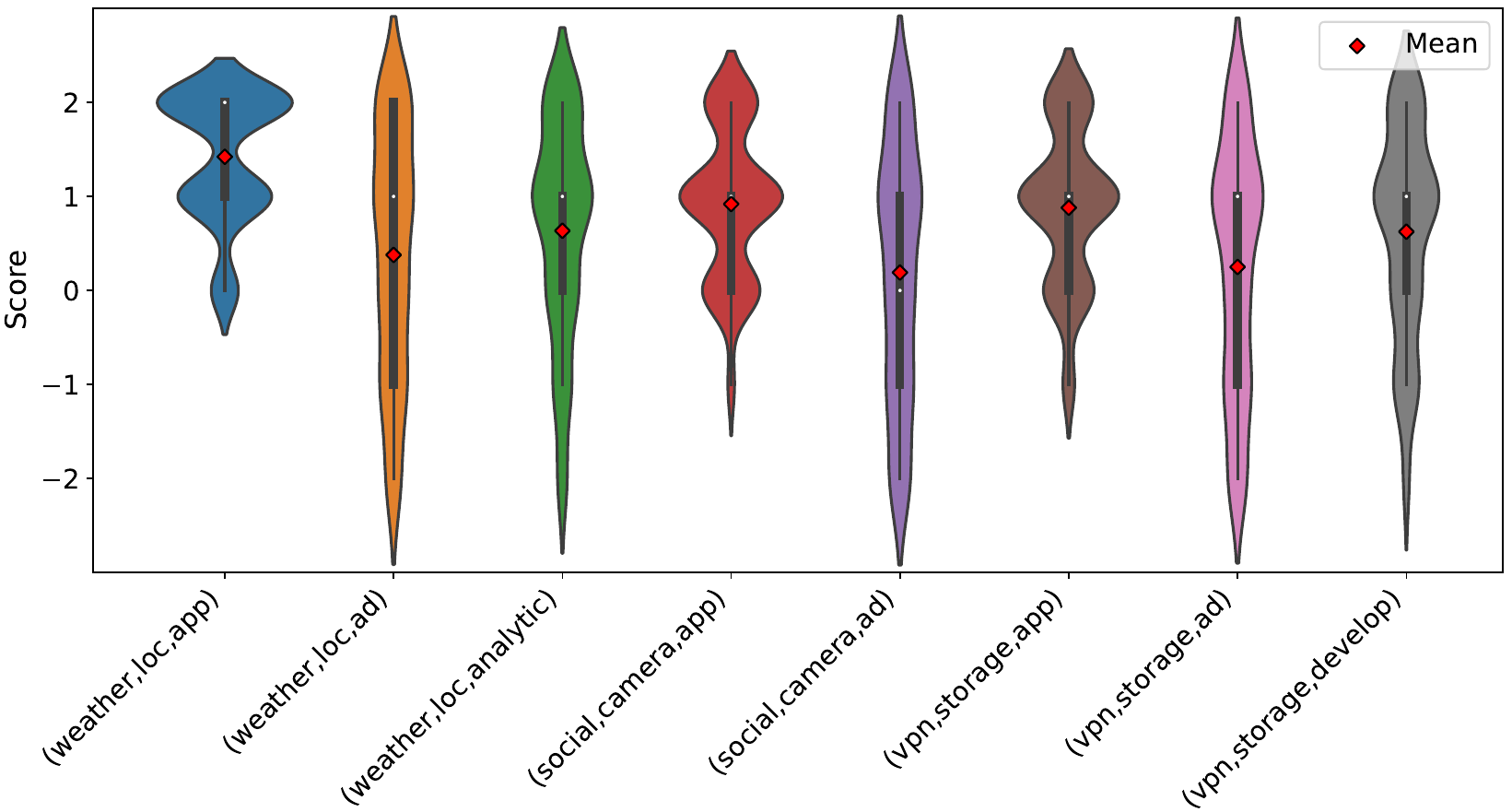}
    \caption{Distribution of users' privacy ratings per question.}
    \label{fig:distribution-1}
\end{figure}

\noindent\textbf{Distribution of Behavior-level Scores.}
Each assessment question corresponds to a specific privacy-related behavior exhibited by a selected app, with participants assigning ratings on a scale from -2 to +2. \autoref{fig:distribution-1} presents a sample of the distribution of these ratings, where the width of each violin indicates the density of ratings at different score levels, with wider regions indicating higher concentrations of user responses. Each question is represented by a triple $\langle app,\ data,\ purpose\_type \rangle$, where $purpose\_type$ denotes either first-party use (\textit{app}, meaning data accessed by the app's own code rather than an embedded third-party SDK) or a third-party SDK category (e.g., \textit{ads}, \textit{analytics}, \textit{develop} for development aid SDKs; see Appendix~\ref{sec:tables} \autoref{tab:sdk_types} for the full list). 
For example, $\langle weather, loc, app \rangle$ signifies that a weather app's own code accesses location data, with the purpose explanation clarifying that this is for delivering localized weather forecasts. We can observe that even when an app accesses the same type of data, the ratings can vary significantly depending on the purpose of use. Generally, participants demonstrate greater acceptance when data access is directly aligned with an app's primary function, but they give more diverse and polarized responses when the same data is utilized for third-party purposes such as advertising, analytics, etc. 
A Mann-Whitney U test confirmed that first-party ratings were significantly higher than third-party ratings ($U = 32{,}986{,}602.5$, $p < .001$, $r = 0.165$), with a small-to-medium effect size indicating that while the difference is consistent and statistically robust, users do not categorically reject third-party data uses; rather, their responses become more varied and context-dependent, reflecting nuanced rather than blanket attitudes.
This suggests that users care more about how and by whom data is used than about the raw data type itself.

\noindent\textbf{Rating Scores for Each App.}
Since user ratings were collected at the level of individual privacy-related behaviors, we next computed an overall privacy rating for each app. The final score, denoted as $AppScore$, was computed based on the following rules: (1) If any negative ratings were present, the final score was determined as the sum of all negative ratings, formulated as 
\begin{equation}
AppScore = \sum s_i,  s_i < 0. 
\end{equation}
This design reflects the asymmetric nature of privacy assessment: a single behavior perceived as seriously inappropriate should not be neutralized by a majority of acceptable ones, analogous to the principle in security evaluation that any critical violation warrants attention regardless of overall compliance. Summing negative ratings also allows differentiation between apps with multiple privacy concerns versus those with only minor issues. (2) If all ratings were non-negative, the final score was computed as the arithmetic mean of all scores for that app, expressed as
\begin{equation}
AppScore = \frac{1}{n} \sum_{i=1}^{n} s_i, \quad s_i \geq 0
\end{equation}
where $n$ represents the number of responses, maintaining a normalized representation when no negative scores are present. We acknowledge that this is one possible aggregation scheme among several alternatives (e.g., weighted sums), and that comparing scores across the two cases involves different scales; exploring alternative aggregation methods remains a direction for future work.

\begin{figure}
    \centering
    \includegraphics[width=0.5\textwidth]{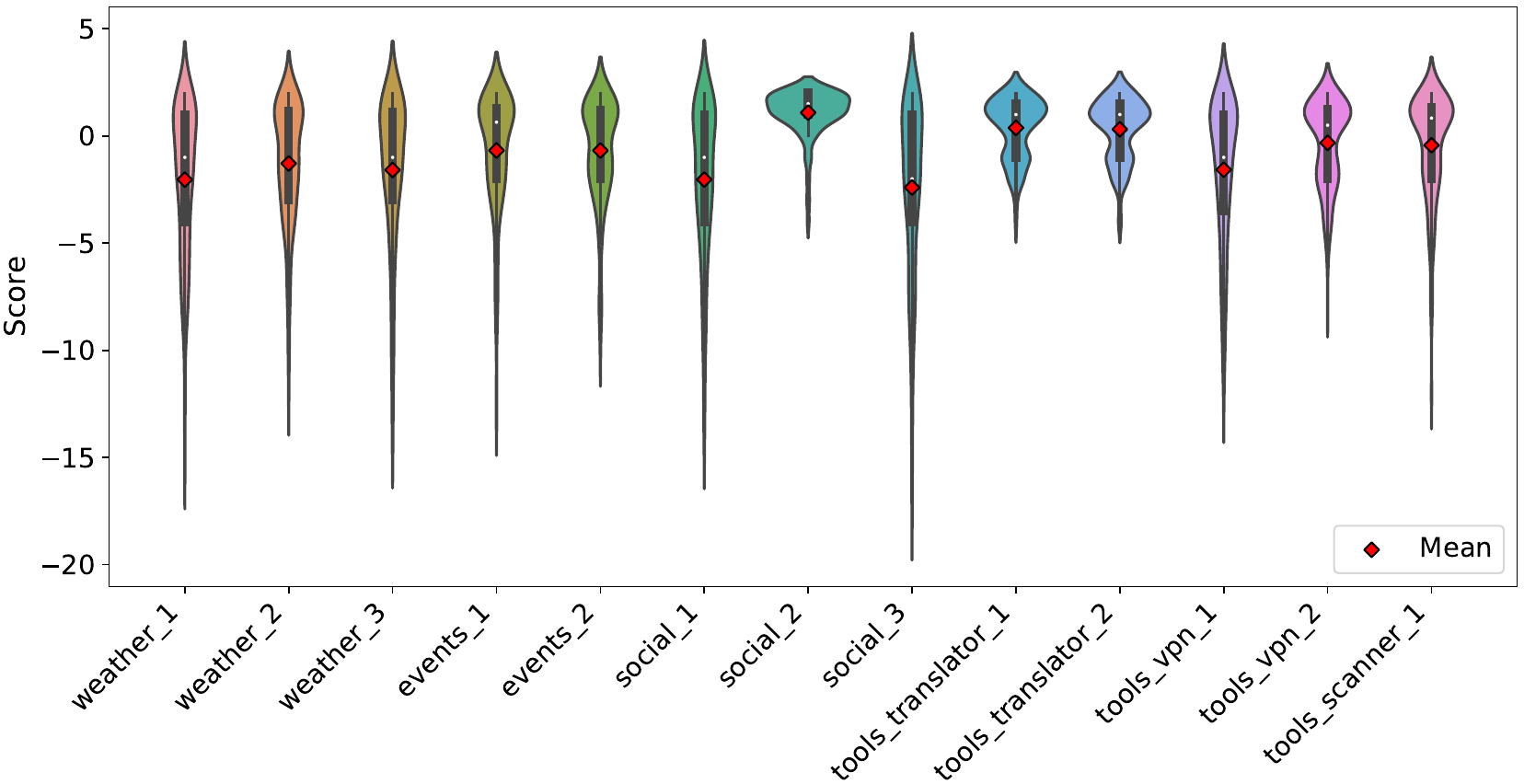}
    \caption{Distribution of users' privacy ratings for each app.}
    \label{fig:distribution}
\end{figure}

\noindent\textbf{Distribution of App-level Scores.}
As shown in \autoref{fig:distribution},  
the visualization captures not only the central tendency of ratings but also the spread and variability of user opinions for each app.
For example, the app \textit{social\_2} exhibits a concentration of positive ratings, indicating that most users found its privacy practices acceptable. The co-presence of positive and negative scores for some apps reflects users' diverse perceptions. These patterns can provide valuable insights for various stakeholders, helping them better understand user sentiment. 
Compared to a single-dimensional score for each app, this approach provides a higher-dimensional representation that captures the full distribution of user ratings.

\noindent\textbf{Summary.}
These findings demonstrate that \system is effective in enabling everyday users to produce informative privacy ratings. At the behavior level, ratings vary meaningfully across different data access scenarios and reflect sensitivity to contextual factors such as purpose of use and data controller–users do not respond uniformly, but differentiate between scenarios in ways that reveal genuine privacy attitudes. At the app level, behavior-level ratings are aggregated into overall app scores that reflect the collective weight of user concerns, giving stakeholders a holistic view of each app's privacy standing from the user perspective. These results confirm the feasibility of the democratized privacy assessment paradigm: ordinary users, when supported by \system, can contribute meaningful and varied privacy evaluations at scale.

\subsection{RQ2: User vs. Expert Ratings}

\subsubsection{Method} 
We asked a panel of four privacy experts to individually evaluate the same set of privacy-related behaviors that had been rated by end-users (cf. RQ1). Each behavior was presented with the same contextual information as in the user study. Experts rated the necessity of the data access (-2 unnecessary ... +2 necessary) based on their expertise and compliance with data minimization principles. 
Each group rated from their natural perspective: users assessed perceived acceptability, while experts assessed regulatory necessity, reflecting how each group would naturally evaluate data access in practice.
We acknowledge that four experts constitute a small panel; the goal of this comparison is not to establish a definitive expert benchmark, but to offer an illustrative contrast between regulatory-informed assessments and everyday user perspectives, thereby exploring where the two groups converge and where they diverge.

\subsubsection{Results} 

We present the results in two parts: expert panel agreement, followed by a comparison between user and expert ratings including both alignments and divergences.
\noindent\textbf{Expert Panel Agreement.}
Krippendorff's $\alpha$ on a representative subset of behavior items was 0.727 (ordinal), indicating substantial inter-rater reliability and exceeding the threshold for tentative conclusions~\cite{krippendorff2011computing}. This level of agreement is consistent with the subjective nature of privacy necessity judgments and itself reveals a meaningful pattern: within the expert group, there was high agreement on clear cases but also divergence in borderline ones. For example, in third-party data usage, an expert contextualized analytics or embedded development integrations as industry norms, whereas others insisted on strict minimization principles.
This shows that expert ratings are not absolute and can vary depending on individual interpretations of compliance standards and personal thresholds for acceptable risk.

\noindent\textbf{User-Expert Alignment.}
Turning to the comparison between users and experts, a Spearman correlation between per-item mean user ratings and mean expert ratings yielded $\rho = 0.737$ ($p < .001$), indicating strong rank-order agreement on the relative acceptability of different behaviors. Both groups, for instance, rated behaviors directly tied to core app functionality—such as weather apps using location for localized forecasts and translator apps accessing the microphone for speech input—as acceptable.
However, users were systematically more lenient in absolute terms: the overall mean user rating was $+0.63$ compared to $-0.28$ for experts, a difference of nearly one full scale point. This gap was most pronounced for third-party data uses (user mean $+0.37$ vs. expert mean $-1.57$): experts rated advertising and analytics-related behaviors near the lowest end of the scale, reflecting strict data minimization standards, while users remained moderately accepting, possibly reflecting habituation to common industry practices.
Conversely, for certain first-party sensitive behaviors such as microphone or camera access for core features, experts rated these as clearly necessary (mean $+2.00$), whereas users were more skeptical, assigning lower but still positive ratings. This suggests that users are less certain than experts about the necessity of sensitive hardware access even when it serves core functionality, highlighting a gap in user awareness that contextual explanations alone may not fully close.

\noindent\textbf{Summary.}
Overall, these findings demonstrate the complementary value of user-generated privacy ratings. The strong rank-order correlation shows that users and experts largely agree on which behaviors are more or less acceptable. At the same time, the divergences between the two groups reflect genuine differences in evaluative perspective rather than noise, and surfacing these differences is precisely what a democratized assessment approach is designed to contribute.

\subsection{RQ3: Risk Preference Based Calibration}
\label{sec:rq3}

\subsubsection{Method}
 
We explored how our proposed risk preference-based adjustment method works in practice. We conducted a post-study survey consisting of four risk-taking scenarios designed to capture participants’ attitudes toward risk in diverse decision-making environments. 
Drawing from prospect theory, the scenarios covered high-probability outcomes with large stakes, low-probability outcomes with small stakes, and equivalent conditions framed as potential losses. For example, participants were asked to choose between a guaranteed outcome and a probabilistic higher outcome in both reward and penalty contexts.
Responses revealed participants’ general preferences in risk, allowing us to contextualize their privacy risk ratings and identify potential biases.
Based on the responses, we classified participants into three groups: 92 (46\%) \textit{risk-averse} and 28 (14\%) \textit{risk-seeking}, and 80 (40\%) \textit{risk-neutral}.
Given this distribution, we set the risk aversion factor to $\lambda=0.6$, slightly above the midpoint to reflect the prevalence of risk-averse tendencies, and tried a moderate adjustment coefficient of $\delta=0.5$.

\subsubsection{Results}

\begin{figure}
    \centering
    \includegraphics[width=0.5\textwidth]{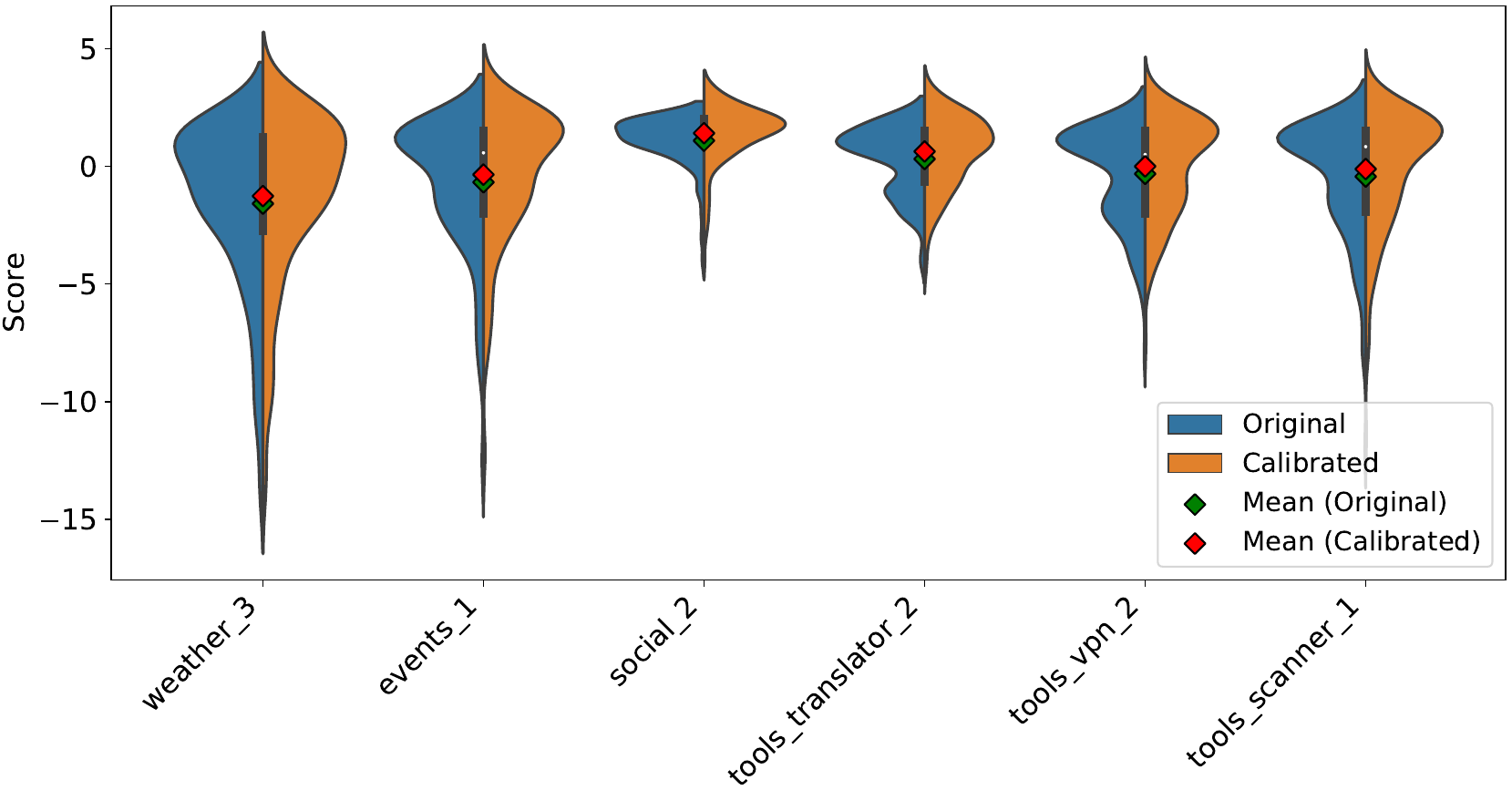}
    \caption{Distribution of users' privacy ratings (original vs. calibrated).}
    \label{fig:calibrated}
\end{figure}
\autoref{fig:calibrated} presents a comparison of original and calibrated ratings across different apps (one per category for visual clarity). Overall, the calibration produced a net upward shift in aggregate scores, driven by the larger proportion of risk-averse participants in our sample, while leaving the relative ordering and overall direction of app assessments intact. We report the effects at three levels of granularity: the overall score distribution, the per-group adjustment patterns, and the per-app outcomes.

\noindent\textbf{Overall Shift.}
The mean aggregate app score increased from $-0.866$ to $-0.547$, a shift of $+0.319$. Notably, 10 out of 13 apps retained negative aggregate scores both before and after calibration, and no app changed its overall valence (i.e., from net-negative to net-positive or vice versa), indicating that calibration refines the magnitude of scores rather than overturning the substantive conclusions of the raw ratings.

\noindent\textbf{Per-Group Effects.}
The asymmetry between risk groups is reflected in the per-group adjustment magnitudes. The 92 risk-averse users, whose original mean score was $-1.048$, had their ratings adjusted upward to $-0.206$ on average (change of $+0.842$). In contrast, the 30 risk-seeking users, whose original mean was $-0.409$, had their ratings adjusted downward to $-0.852$ (change of $-0.443$). The larger upward adjustment for risk-averse users compared to the downward adjustment for risk-seeking users reflects both the higher proportion of risk-averse participants and the greater deviation of their scores from the neutral baseline. The 80 risk-neutral users remained unchanged. Together, these adjustments pull the aggregate distribution toward a more balanced representation of the overall user population.

\noindent\textbf{Per-App Patterns.}
At the app level, the three apps that received positive aggregate scores before calibration (\textit{social\_2}, \textit{tools\_translator\_1}, and \textit{tools\_translator\_2}) remained positive after calibration, with scores increasing modestly (e.g., \textit{social\_2}: $+1.088 \to +1.404$). Among the negatively-rated apps, \textit{tools\_vpn\_2} showed the most notable movement, approaching the neutral point ($-0.320 \to -0.004$), while \textit{social\_3} and \textit{weather\_1} remained the most negatively rated apps even after calibration ($-2.082$ and $-1.716$ respectively), suggesting strong and robust user concern about these apps' data practices regardless of risk-preference correction.

\noindent\textbf{Summary.}
These results demonstrate that individual risk preferences systematically influence privacy ratings, and that the proposed adjustment mechanism can account for this variation. By identifying the risk profile of participants and calibrating their ratings accordingly, \system produces aggregate scores that better reflect the general user population rather than being skewed by the disproportionate representation of risk-averse individuals. This shows the feasibility of incorporating behavioral science-informed adjustments into democratized privacy evaluation.

%% file: discussion.tex
\section{Discussion}

\subsection{Design Implications}
Our study yields several actionable insights for designing user-grounded privacy assessment systems, as well as broader implications for privacy research and HCI.

\noindent\textbf{Contextual framing is the primary unit of privacy judgment.}
Both the participatory design workshop and the large-scale rating study show that users rarely base their judgments on permissions alone; instead, they reason about whether specific data accesses are proportionate to an app’s core functionality and stated purpose.
This pattern aligns with contextual integrity theory~\cite{nissenbaum2004privacy} and suggests that it operates not only as a theoretical framework but as a genuine cognitive heuristic that everyday users apply.
For system designers, this reinforces that purpose explanations and functionality-based framing should be treated as primary interface elements rather than supplementary disclosures.
For privacy researchers, it suggests that permission-centric studies may systematically underestimate users’ capacity for nuanced judgment when appropriate context is provided.

\noindent\textbf{Disclosure quality matters more than disclosure quantity.}
Participants expressed a desire for clear explanations but also reported feeling overwhelmed by verbose details.
This finding moves the conversation in usable privacy research~\cite{lin2012expectation,ismail2015crowdsourced} beyond the question of \textit{whether} to disclose toward \textit{how much} to disclose, and in what structure.
Effective designs should prioritize the most decision-relevant cues upfront while keeping additional context accessible in a structured, navigable form.
The two-tier format adopted in \system (concise header plus contextual explanation) operationalizes this principle; future work could evaluate whether layered disclosure formats consistently reduce cognitive load without sacrificing informed decision-making.

\noindent\textbf{User-expert divergences reflect legitimate evaluative pluralism, not measurement noise.}
A key finding is that user and expert ratings are meaningfully correlated ($\rho = 0.737$, $p < .001$) yet exhibit systematic divergences.
These divergences are not random: they reflect structurally different evaluative frames–users reason primarily from perceived acceptability and contextual appropriateness, while experts apply regulatory necessity criteria.
This challenges the implicit assumption in privacy auditing practice that expert assessments represent a singular ground truth against which user opinions should be calibrated.
Our findings suggest instead that user and expert perspectives constitute complementary forms of evidence; privacy governance frameworks that incorporate both can capture a fuller picture of what constitutes appropriate data access in practice.

\noindent\textbf{Aggregate privacy scores implicitly encode population assumptions.}
The risk-preference calibration analysis reveals that aggregate privacy scores are sensitive to the composition of the evaluating population: a sample skewed toward risk-averse users will produce more negative aggregate scores than a balanced one, even when rating the same apps.
This has methodological implications beyond \system: any privacy metric derived from user judgments carries an implicit assumption about the underlying population’s risk distribution.
Making this assumption explicit and designing scoring pipelines that account for it is a prerequisite for privacy assessments that are both representative and reproducible.
Our calibration approach, which adjusts ratings based on measured risk preferences before aggregation, offers one concrete way to operationalize this principle.

\subsection{Visions for Advancing the Paradigm}
\system is still in its infancy, but its underlying conceptualization and rationale have promising potentials. 
We acknowledge that recruiting 200 participants via Prolific constitutes a proof-of-concept demonstration rather than a fully realized democratization of privacy evaluation. The term ``democratizing'' in this work refers to broadening the scope of privacy evaluation by actively incorporating everyday user perspectives as a complement to expert assessments, and its full realization requires large-scale, real-world deployment.

\noindent\textbf{Integration into App Ecosystems.}
A significant step forward would be embedding the system within major app ecosystems like Google Play and App Store. This integration would enable privacy ratings to become as commonplace as user reviews, allowing consumers to make privacy-conscious choices while holding developers accountable for their data practices. 
Prior work has shown that user privacy reviews can prompt developers to reduce their apps' permission usage~\cite{nguyen2019short}, and existing platform mechanisms (e.g., privacy nudges in the Google Play developer console) have demonstrated the effectiveness of feedback-driven interventions. \system could extend such mechanisms by incorporating structured, aggregated user privacy ratings, providing developers with richer and more actionable signals than unstructured reviews alone.

\noindent\textbf{AI-Enhanced Privacy Evaluation.}
As user participation grows, AI models trained on aggregated user ratings could help automate the detection of privacy-problematic patterns at scale, extending coverage beyond what either manual expert audits or user ratings alone can sustain.
A particularly promising direction is the development of personalized user-role agents that simulate diverse privacy preferences and risk attitudes, enabling privacy evaluations to be tailored to specific user populations rather than relying on a single aggregate standard.
These directions point toward hybrid human-AI pipelines that preserve the human-grounded legitimacy of democratized evaluation while overcoming its scalability constraints.

\subsection{Limitations and Future Work}

\noindent \textbf{Participant Selection.} 
The participant sample may not fully represent the broader mobile user population. While we aimed for diversity in our participatory design and Prolific-based study, participants could still skew toward higher-than-average privacy awareness.
Moreover, the expert panel comprised only four professionals, which limits the representativeness of the expert-side comparison in RQ2. A larger and more diverse panel spanning regulatory, legal, and technical backgrounds would be needed to establish a more robust expert baseline.

\noindent \textbf{Response Biases.} 
As with many survey-based studies, our design may be subject to social desirability and fatigue effects. 
For example, the relatively high proportion of participants claiming to “often” read privacy policies may reflect socially desirable responding rather than actual behavior. 
Question wording may also have contributed, as terms like ``setting permissions'' and ``privacy policies'' are open to broader interpretations that could inflate self-reported familiarity rates.
Answering 78 assessment items in one session could also cause mild fatigue, potentially affecting attention or consistency.
We mitigated these effects through attention checks and pilot testing; however, future work could further reduce them via shorter, adaptive tasks or behavioral validation.

\noindent \textbf{Proof-of-Concept Technical Pipeline.}
The automated pipeline for purpose explanation generation and first/third-party classification is a prototype-level implementation, not a production-ready solution. 
The pipeline draws on a limited set of information sources (static API analysis and app store descriptions) omitting complementary signals such as the Data Safety Section on Google Play, GUI-level permission rationale messages, privacy policy text, and SDK-level opt-in/opt-out disclosures. These limitations do not undermine the validity of our user study, as all 78 deployed items were individually verified by the research team prior to deployment to ensure accuracy. Improving controller classification and incorporating richer information sources remain important directions for future work.

\noindent \textbf{Simplified Modeling of User Preferences.}
Our current approach represents an initial attempt to account for user heterogeneity by applying prospect theory and collecting risk preference data through a simple survey. While it serves the purpose of the proof of concept well, it could be further enhanced. Future work could explore more sophisticated techniques, such as behavioral modeling, longitudinal profiling, or context-aware personalization, to capture users’ privacy preferences in a richer and more dynamic manner.

\noindent \textbf{System Usability.}
The current evaluation focused on the quality of rating outputs rather than a formal assessment of \system's usability. Although high task completion rates and low attention-check failure rates suggest the system was accessible in practice, future work should include structured usability evaluations such as think-aloud protocols or standardized questionnaires to more rigorously assess whether specific design features (e.g., contextual explanations, the rating interface) effectively support users in making informed privacy judgments.

\noindent \textbf{Concern on Potential Abuse.} Some users from PD pointed out a potential concern for democratized privacy ratings: the risk of manipulation or abuse, similar to the “rating inflation” or “review bombing” observed in app stores. 
While not addressed in the current version of \system (as it is better suited as a later-stage safeguard), it does not undermine the value of the paradigm itself. Rather, it calls for complementary safeguards of the paradigm, such as anomaly detection and user reputation mechanisms, 
to prevent potential abuse of the system in future work.

%% file: relatedwork.tex
\section{Related Work}
\noindent \textbf{Privacy Risk Assessments for Mobile Apps.}
Regulatory guidelines have driven extensive research into automated privacy risk assessments for mobile apps~\cite{chang2020framework,kim2023privacy,zhou2023policycomp,tang2017detecting,yang2021pradroid,kang2015visualizing}.  
For example,
Chang et al.~\cite{chang2020framework} developed an automated framework that estimates privacy risks by analyzing app permissions and privacy policies using probabilistic models. Kim et al.~\cite{kim2023privacy} proposed six key indicators  (Ability, Benevolence, Integrity, Experience, Reputation, and Inclination) to quantify privacy scores in the context of handling personal data. 
Yang et al. \cite{yang2021pradroid} applied information flow analysis to construct a risk matrix, correlating permissions with potential severity levels to assign a privacy risk score to apps.
These studies present solid approaches to evaluating privacy risks, but they largely rely on professional knowledge and understanding. 
Our work complements them by incorporating end-users’ perspectives into the assessment, extending the scope of evaluation beyond expert-centric models.

\noindent \textbf{User Perspectives on Mobile Privacy.}
Recent research~\cite{lin2012expectation,lin2014modeling,ismail2015crowdsourced,bongard2022necessary,hamed2016privacy,kelley2013privacy} has highlighted the importance of user mental models in shaping their privacy judgments and trust in mobile apps.
Lin et al.~\cite{lin2012expectation} were among the first to use crowdsourcing to analyze users' mental models of mobile app behavior, demonstrating that users' expectations regarding data access significantly influence trust decisions.
Ismail et al.~\cite{ismail2015crowdsourced} conducted a user study to determine the minimal set of permissions required for app usability across diverse users, paving the way for user-driven privacy evaluations.
These studies enlighten that end-users can contribute valuable, context-aware insights that complement automated tools and regulatory audits. 
Building on this line of research, we take a step toward a democratizing privacy assessment paradigm that moves beyond eliciting user perceptions and operationalizes them within a structured, scalable, and actionable evaluation framework. This paradigm repositions users as active evaluators in the privacy auditing process and explores methods to actively engage users in assessing the appropriateness of apps' privacy-related behaviors, thereby bridging the gap between user-centric insights and systematic privacy evaluation practices.

%% file: conclusion.tex
\section{Conclusion}

This paper proposes a novel paradigm for the democratization of privacy assessment and introduces \system, a prototype infrastructure that empowers everyday users to evaluate mobile apps' data access behaviors based on their perceived necessity and appropriateness. Grounded in participatory design with everyday users, \system incorporates four key features: contextual explanation provision, category-based representative selection, an intuitive rating interface, and preference-based rating adjustment.
Our evaluation with 200 participants demonstrates that everyday users can provide meaningful privacy ratings that both align with and diverge from expert assessments in informative ways. These divergences are not noise but signal: they reveal dimensions of privacy concern, particularly around contextual appropriateness and data minimization, that complement the legal and technical focus of expert-driven evaluations. 
Beyond the specific findings, this work offers the HCI community a set of design implications grounded in both participatory design and large-scale user evaluation. These implications around surfacing contextual justifications, balancing transparency with cognitive load, and accounting for heterogeneous risk preferences provide actionable guidance for future privacy tools and assessment frameworks.
We envision \system as a foundation for more transparent, participatory, and user-oriented privacy governance that complements expert-driven oversight and fosters accountability among app developers at scale.

%% file: appendix.tex
\clearpage
\appendix
\label{sec:appendix}

\section*{Supplementary Tables}
\label{sec:tables}

\autoref{tab:data_permissions} presents the set of high-impact data types and their associated Android permissions considered in our study. These include core resources such as calendars, call logs, contacts, location, microphone, sensors, SMS, storage, and others, which represent sensitive access points frequently highlighted in privacy research and regulatory discussions.
By explicitly mapping each data type to its relevant permissions, the table provides a concrete scope of the privacy-sensitive behaviors under investigation, forming the foundation for both our system’s contextual explanations and the user evaluations of data access practices.

\begin{table}[htbp]
    \centering
    \caption{The list of our concerned data/resource types and their corresponding permissions.}
    \resizebox{\linewidth}{!}{%
    \begin{tabular}{ll}
        \toprule
        \textbf{Data Type} & \textbf{Permissions} \\
        \midrule
        CALENDAR & READ\_CALENDAR \\
                 & WRITE\_CALENDAR \\
        \midrule
        CALL\_LOG & READ\_CALL\_LOG \\
                  & WRITE\_CALL\_LOG \\
        \midrule
        CAMERA & CAMERA \\
        \midrule
        CONTACTS & READ\_CONTACTS \\
                 & WRITE\_CONTACTS \\
                 & GET\_ACCOUNTS \\
        \midrule
        LOCATION & ACCESS\_FINE\_LOCATION \\
                 & ACCESS\_COARSE\_LOCATION \\
                 & ACCESS\_BACKGROUND\_LOCATION \\
        \midrule
        MICROPHONE & RECORD\_AUDIO \\
        \midrule
        PHONE & READ\_PHONE\_STATE \\
              & READ\_PHONE\_NUMBERS \\
        \midrule
        SENSORS & BODY\_SENSORS \\
                & BODY\_SENSORS\_BACKGROUND \\
        \midrule
        ACTIVITY\_RECOGNITION & ACTIVITY\_RECOGNITION \\
        \midrule
        SMS & SEND\_SMS \\
            & RECEIVE\_SMS \\
            & READ\_SMS \\
            & RECEIVE\_WAP\_PUSH \\
            & RECEIVE\_MMS \\
            & READ\_CELL\_BROADCASTS \\
        \midrule
        STORAGE & READ\_EXTERNAL\_STORAGE \\
                & WRITE\_EXTERNAL\_STORAGE \\
                & MANAGE\_EXTERNAL\_STORAGE \\
                & ACCESS\_MEDIA\_LOCATION \\
                & READ\_MEDIA\_IMAGES \\
                & READ\_MEDIA\_VIDEO \\
                & READ\_MEDIA\_AUDIO \\
                & READ\_MEDIA\_VISUAL\_USER\_SELECTED \\
        \bottomrule
    \end{tabular}%
    }
    \label{tab:data_permissions}
\end{table}

\autoref{tab:sdk_types} presents ten categories of mobile third-party SDKs, grouped according to their primary functionalities. The categories cover a wide spectrum of services, including development utilities, advertising, analytics, mapping, payment processing, social networking, GUI components, game engines, digital identity, and app marketplaces. Each category reflects common forms of third-party integrations in mobile apps, many of which involve sensitive data collection (e.g., advertising and analytics tracking user behavior, digital identity managing authentication data). By organizing third-party services into these functional groups, we provide a clearer view of how external SDKs contribute to app functionality while also shaping privacy risks, offering a systematic basis for analyzing third-party data access and informing user-facing privacy explanations.

\begin{table}[htbp]
    \centering
    \caption{Types of mobile third-party SDKs}
    \resizebox{\linewidth}{!}{%
    \begin{tabular}{lp{10cm}}
        \toprule
        \textbf{SDK Type} & \textbf{Description} \\
        \midrule
        Development Aid & Libraries that provide development utilities such as parsers, debugging tools, and code analysis utilities. \\
        \midrule
        Advertisement & SDKs used for integrating in-app advertisements, enabling revenue generation through ad impressions and clicks. \\
        \midrule
        Mobile Analytics & SDKs that collect and analyze user data, tracking user engagement, retention, and performance metrics. \\
        \midrule
        Map & SDKs that provide mapping, navigation, and geolocation services to enhance location-based functionalities in applications. \\
        \midrule
        Payment & SDKs that enable in-app payment processing, supporting various payment gateways and digital wallet integrations. \\
        \midrule
        Social Network & SDKs that integrate social media functionalities, such as sharing, authentication, and social graph interactions. \\
        \midrule
        GUI Component & Libraries providing reusable UI elements, such as widgets, animations, and design frameworks for applications. \\
        \midrule
        Game Engine & SDKs that provide a framework for game development, including physics engines, rendering, and asset management. \\
        \midrule
        Digital Identity & SDKs used for authentication and identity verification, supporting OAuth, biometric authentication, and SSO. \\
        \midrule
        App Market & SDKs that facilitate app distribution, updates, and user acquisition through third-party app stores and marketplaces. \\
        \bottomrule
    \end{tabular}%
    }
    \label{tab:sdk_types}
\end{table}

\clearpage
\section*{Post-Study Survey}
\label{sec:survey}
\subsubsection*{1. Background Information}

\begin{itemize}
    \item Prolific ID: \underline{\hspace{2cm}} (Required)
    \item Daily time spent using mobile apps: \underline{\hspace{1cm}} (Required)
\end{itemize}

Types of apps you frequently use: (Select at least one)\\
\begin{tabular}{ll}
    $\square$ Social & $\square$ Shopping \\
    $\square$ Weather & $\square$ Business \\
    $\square$ Education & $\square$ Maps and Navigation \\
    $\square$ Music and Audio & $\square$ Health and Fitness \\
    $\square$ Others & 
\end{tabular}

\subsubsection*{2. Mobile Privacy Preferences}
\mbox{}\\[0.5em]
\quad\textbf{A. Sensitivity of Data Types}\\
Which of the following user data types do you consider sensitive? (Select at least one)\\
\begin{tabular}{ll}
    $\square$ Name and Contact Information & $\square$ Geographic Location \\
    $\square$ Call Records & $\square$ App Usage Records \\
    $\square$ Biometric Data & $\square$ Health Data \\
    $\square$ Bank Transaction Data & $\square$ Photos
\end{tabular}

Other sensitive data types (if any): \underline{\hspace{4cm}}

\textbf{B. Awareness of Privacy Rights}\\
Do you carefully read privacy policies when installing apps? (Required)\\
$\bigcirc$ Often \quad $\bigcirc$ Sometimes \quad $\bigcirc$ Rarely

How familiar are you with setting app permissions? (Required)\\
$\bigcirc$ Very Familiar \quad $\bigcirc$ Somewhat Familiar \quad $\bigcirc$ Not Familiar

\textbf{C. App Behavior Choices}\\
If you discover an app might leak privacy data, what actions would you take? (Select at least one)\\
\begin{tabular}{l}
    $\square$ Stop using the app\\ 
    $\square$ Look for alternative apps \\
    $\square$ Continue using but restrict permissions \\ $\square$ Contact the developer or relevant authorities
\end{tabular}

Other actions you might take (if any): \underline{\hspace{4cm}}

\subsubsection*{3. Risk Attitude Test}
\mbox{}\\[0.5em]
\quad\textbf{1. You participated in a lottery. Which option would you choose?} (Required)\\
$\bigcirc$ A. Guaranteed \$90 reward \\ $\bigcirc$ B. 95\% chance to win \$100, 5\% chance to win nothing

\textbf{2. You participated in a lucky draw. Which option would you choose?} (Required)\\
$\bigcirc$ A. Guaranteed \$5 reward \\ $\bigcirc$ B. 5\% chance to win \$100, 95\% chance to win nothing

\textbf{3. You need to pay a fee to participate in a game. Which option would you choose?} (Required)\\
$\bigcirc$ A. Pay \$90 for sure \\ $\bigcirc$ B. 95\% chance to pay \$100, 5\% chance to pay nothing

\textbf{4. You were told you might need to pay a fine. Which option would you choose?} (Required)\\
$\bigcirc$ A. Pay \$5 for sure \\ $\bigcirc$ B. 5\% chance to pay \$100, 95\% chance to pay nothing

\subsubsection*{4. User Preferences and Attitudes}

\quad\textbf{Which type of app do you prefer?} (Required)\\
$\bigcirc$ Free apps with extensive permissions \\ $\bigcirc$ Paid apps without personal data collection

What are your thoughts on mobile privacy protection?\\
\underline{\hspace{8cm}}\\
\underline{\hspace{8cm}}
